\documentclass[preprint2]{proto}
\usepackage{times}
\newcommand{\refs}{\par\noindent\hangindent=1pc\hangafter=1}
\def\h{H$_2$}
\def\kms{km\,s$^{-1}$}
\def\deg{{$^\circ$}}
\def\sun{$_\odot$}
\def\Msun{M$_\odot$}

\def\Lsun{L$_\odot$}
\def\Msunyr{M$_\odot$\,yr$^{-1}$}
\def\Msunkmsyr{M$_\odot$\,km\,s$^{-1}$yr$^{-1}$}
\def\water{H$_2$O}
\def\gtsim{\lower.5ex\hbox{$\buildrel > \over\sim$}}
\def\ltsim{\lower.5ex\hbox{$\buildrel < \over\sim$}}

\newcommand{\twco}{$^{12}$CO \/}

\begin{document}

\title{\textbf{\LARGE Molecular Outflows in Low- and High-Mass Star Forming Regions}}

\author {\textbf{\large H\'ector G. Arce}}
\affil{\small\em American Museum of Natural History}
\author {\textbf{\large Debra Shepherd}}
\affil{\small\em National Radio Astronomy
  Observatory}
\author {\textbf{\large Fr\'ed\'eric Gueth}}
\affil{\small\em Institut de Radioastronomie Millim\'etrique, Grenoble}
\author {\textbf{\large Chin-Fei Lee}}
\affil{\small\em Harvard-Smithsonian Center for Astrophysics}
\author {\textbf{\large Rafael Bachiller}}
\affil{\small\em Observatorio Astron\'omico Nacional}
\author {\textbf{\large Alexander Rosen}}
\affil{\small\em  Max-Planck-Institut f\"ur Radioastronomie}
\author {\textbf{\large Henrik Beuther}}
\affil{\small\em Max-Planck-Institut f\"ur Astronomie}

\begin{abstract}
\baselineskip = 11pt
\leftskip = 0.65in 
\rightskip = 0.65in
\parindent=1pc

  We review the known properties of molecular outflows from 
  low- and high-mass young stars. General
  trends among outflows are identified, and the most recent studies on
  the morphology, kinematics, energetics, and evolution of molecular
  outflows are discussed, focusing on results from high-resolution
  millimeter observations.  We review the existing four broad
  classes of outflow models and compare numerical simulations with the
  observational data.  A single class of models cannot explain the
  range of morphological and  kinematic properties that are observed,
  and we propose a possible solution. 
   The impact of outflows on their
  cloud is examined, and we review how outflows can disrupt their
  surrounding environment, through the clearing of gas and the
  injection of momentum and energy onto the gas at distances from
  their powering sources from about 0.01 to a few pc.  We also
  discuss the effects of shock-induced chemical processes on the
  ambient medium, and how these processes may act as a chemical clock
  to date outflows.  Lastly, future outflow research with existing and
  planned millimeter and submillimeter instruments is presented.
  \\~\\~\\~
\end{abstract}  

\section{\textbf{INTRODUCTION}}

As a star forms by gravitational infall, it energetically expels mass
in a bipolar jet.  There is strong evidence for a physical link
between inflow and outflow and that magnetic stresses in the 
circumstellar disk-protostar system
initially launch the outflowing material 
(see chapters by {\it Pudritz et al.}; 
{\it Ray et al.}; and {\it Shang et al.}).  
The ejected matter can
accelerate entrained gas to velocities greater than those of the
cloud, thereby creating a molecular outflow. Outflows can induce
changes in the chemical composition of their host cloud and may even
contribute to the decline of the infall process by clearing out dense
gas surrounding the protostar.  In addition, molecular outflows can be useful
tools for understanding the underlying formation process of stars
of all masses, as they provide a record of the mass-loss history of
the system.

Protostellar outflows can be observed over a broad range of
wavelengths, from the ultraviolet to the radio. In this review we will
concentrate on the general characteristics and properties of molecular
outflows, the entrainment process, and the chemical and physical
impact of outflows on the cloud that are mainly detected through
observations of molecular rotational line transitions at millimeter
and submillimeter wavelengths. At these wavelengths the observations
mainly trace the cloud gas that has been swept-up by the underlying
protostellar wind, and provide a time-integrated view of the
protostar's mass-loss process and its interaction with the surrounding
medium.


\section{\textbf{GENERAL OUTFLOW PROPERTIES}}

Over the last 10 years, millimeter interferometers have allowed the
observation of molecular outflows at high angular resolutions ($\sim
1$ to $4"$), while the capability to observe mosaics of several
adjacent fields has enabled mapping of complete outflows at those
resolutions. Such interferometric observations give access to the
internal structure of the gas surrounding protostars, and 
can disentangle the morphology and dynamics of the
different elements that are present (i.e., protostellar condensation,
infalling and outflowing gas).  These high resolution
observations have been critical to the discovery of the kinematics and
morphology of outflows from massive OB (proto)stars, which are
typically more than a kiloparsec away.

General trends have been identified in molecular outflows from both
low- and high-mass protostars, even though they display a broad
diversity of sizes and shapes.  These properties have been identified
mostly using single-dish and interferometer observations of the CO
lines.  Molecular outflows exhibit a mass-velocity relation with a
broken power law appearance, $dM(v)/dv \propto v^{-\gamma}$, with the
slope, $\gamma$, typically ranging from 1 to 3 at low outflow
velocities, and a steeper slope at higher velocities --- with $\gamma$ as
large as 10 in some cases (e.g., 
{\it Rodr\'{\i}guez et al.}, 1982; {\it Lada and
Fich}, 1996; {\it Ridge and Moore}, 2001).  
The slope of the mass-velocity
relation steepens with age and energy in the flow ({\it Richer et
al.}, 2000).  
The velocity at which the slope changes is typically between 6 and 12~{\kms} 
although outflows can have CO break velocities as low as about 2~{\kms} and, in the
youngest CO outflows, it can be high as 
 30~{\kms} (see, e.g., {\it Richer et al.}, 2000, 
and references therein).  
The mass, force, and mechanical luminosity of
molecular outflows correlate with bolometric luminosity 
({\it Bally and Lada},
1983; {\it Cabrit and Bertout}, 1992; {\it Wu et al.}, 2004), and many fairly
collimated outflows show a linear velocity-distance relation,
typically referred to as the ``Hubble-law'', where the maximum radial
velocity is proportional to position (e.g., {\it Lada and Fich}, 1996).  Also,
the degree of collimation of outflows from low- and high-mass systems
appears to decrease as the powering source evolves (see below).

These observed general trends are consistent with a common
outflow/infall mechanism for forming stars with a wide range of masses,
from low-mass protostars up to early B
protostars.  
Although there is evidence that
the energetics for at least some early-B stars may differ from their
low-mass counterparts, the dynamics are still governed by the presence
of linked accretion and outflow.  A few young O stars show evidence
for accretion as well although this is not as well established as for
early-B stars (e.g., {\it van der Tak and Menten}, 2005; chapter by 
{\it Cesaroni et al.}).

\bigskip
\noindent
\textbf{2.1 Outflows from low-mass protostars}
\bigskip

Since their discovery in the early eighties, molecular outflows driven
by young low-mass protostars (i.e.\ typically $<1$~\Msun) have been
extensively studied, giving rise to a detailed picture of these
objects (see, e.g., the reviews by 
{\it Richer et al.},  2000; {\it Bachiller and
Tafalla}, 1999, and references therein). The flows typically extend over
0.1--1 parsec, with outflowing velocities of 10--100~\kms.  Typical
momentum rates of $10^{-5}$~\Msunkmsyr\ are observed, while the    
molecular outflow mass flux can be as high as $10^{-6}$~\Msunyr \/ 
({\it Bontemps et al.}, 1996). 
Particular interest has been
devoted to the outflows driven by the youngest, embedded protostars
(age of a few $10^3$ to a few $10^4$ years, the Class~0 objects). These sources are
still in their main accretion phase and are therefore at the origin of
very powerful ejections of matter.

{\em 2.1.1. Molecular jets}. 
The collimation factor 
(i.e., length/width, or major/minor radius) of the CO outflows, as
derived from single-dish studies, range from $\sim$3 to $>$20. There
is however a clear trend of higher collimation at higher outflowing
velocities (see, e.g., {\it Bachiller and Tafalla}, 1999).  Interferometric
maps have revealed even higher collimation factors, and, in some
cases, high-velocity structures that are so collimated (opening angles
$<$ a few degrees) that they can be described as ``molecular jets''.

HH\,211 is the best example to date of such a molecular jet ({\it Gueth and
Guilloteau}, 1999). At high-velocity, the CO
emission is tracing a highly-collimated linear structure that is
emanating  from the central protostar. This CO jet terminates at the
position of strong H$_2$ bow-shocks, and shows a Hubble law velocity
relation. Low-velocity CO traces a cavity
that is very precisely located in the wake of the shocks. These
observations strongly suggest that the propagation of one or several
shocks in a protostellar jet entrain the ambient
molecular gas and produces the low-velocity molecular outflow (see Sec.~3).  
With an estimated 
dynamical timescale of $\sim$10$^3$ years, HH\,211 is  obviously an
extremely young object. Other examples of such highly-collimated,
high-velocity jets include IRAS\,04166+2706 ({\it Tafalla et al.}, 2004) and
HH\,212 ({\it Lee et al.}, 2000) -- these sources are or will be in the near
future the subject of more detailed investigations.

In at least 
SVS\,13B ({\it Bachiller et al.}, 1998, 2000), and   
NGC1333 IRAS 2 ({\it J\o rgensen et al.}, 2004), NGC1333 IRAS 4
({\it Choi} 2005) 
and HH\,211 ({\it Chandler and Richer}, 2001; 
{\it Hirano et al.}, 2006; {\it Palau et al.}, 2006)
the SiO emission 
traces the molecular jet and {\em not} the strong terminal shocks
against the interstellar medium.  This came as a surprise, as it seems
 to contradict the widely accepted idea that SiO is a 
tracer of outflow shocks, where the density is increased by several
order of magnitudes (e.g., {\it Mart\'{i}n-Pintado et al.}, 1992; 
{\it Schilke et al.}, 1997; {\it Gibb et al.}, 2004).  
The lack of significant SiO emission in the terminal
shocks  suggests that the formation process of this molecule has
a strong dependence on the shock conditions (velocity, density)
and/or outflow age (see Sec.~4.2).

The exact nature of these CO and SiO molecular jets is not yet clear.
Three basic scenarios could be invoked, in which the high-velocity CO
and SiO molecules {\em (a)} belong to the actual protostellar jet,
{\em (b)} are entrained along the jet in a turbulent cocoon (e.g.,
{\it Stahler}, 1994; {\it Raga et al.}, 1995), or {\em (c)} are formed/excited in
shocks that are propagating down the jet (``internal working
surfaces'', {\it Raga and Cabrit}, 1993). This latter scenario would reconcile
the observation of SiO in the jet and the shock-tracer nature of this
molecule.  The predictions of these three cases, both in terms of line
properties and observed morphologies, are somewhat different but the
current  observations have not yet provided a clear preference for
one of these scenarios.

{\em 2.1.2. More complex structures}.
 Not all sources have structures as
simple or unperturbed as the molecular jets discussed
above. CO observations have also revealed a number of more
complex outflow properties.

Episodic ejection events seem to be a common property of young
molecular outflows. In sources such as, e.g., L\,1157 ({\it Gueth et al.},
1998) and IRAS\,04239+2436 (HH\,300, {\it Arce and Goodman}, 2001b), a limited
number (2 to 5) of strong ejection events  have taken place, each
of them resulting in the propagation of a large shock.
Morphologically, the flow is therefore the superposition of several
shocked/outflowing gas structures, while position-velocity diagrams
show multiple ``Hubble wedges'' (i.e., a jagged profile; {\it Arce and
Goodman}, 2001a). In most of the sources, if several strong shocks are
not present, a main ejection event followed by several smaller, weaker
shocked areas are observed (e.g., L1448: {\it Bachiller et al.}, 1990; HH\,111:
{\it Cernicharo et al.}, 1999; several sources: {\it Lee et al.}, 2000, 2002). 
 As noted before, even the molecular jets  could
include several internal shocks. Altogether, these properties suggest
that the ejection  phenomenon in young outflows is intrinsically
episodic, or --- a somewhat more attractive possibility --- could be
continuous but include frequent ejection bursts. This could be
explained by sudden variations in the accretion rate onto the forming
star, that result in variations of the velocity of the 
ejected matter, hence the creation  of a series of shocks.

Precession of the ejection direction
has been established in a
few sources, like  Cep~E
({\it Eisl\"offel et al.}, 1996), and L\,1157 ({\it Gueth et al.}, 1996, 1998).
 In several other objects, the observations
reveal bending or  misalignment between the structures within
the outflows (see e.g., {\it Lee et al.},  2000, 2002). In fact, when
observed at the angular resolution provided by millimeter
interferometers, many well-defined, regular bipolar outflows mapped
with single-dish telescopes often reveal much more complex and 
irregular structures, which indicate both temporal and spatial
variations of the ejection  phenomenon.

Quadrupolar outflows
are sources in which four lobes are
observed, and seem to be driven by the same protostellar
condensation. Several scenarios were proposed to explain these
peculiar objects: two  independent outflows (e.g., {\it Anglada et al.},
1991; {\it Walker et al.},1993); one single flow with strong
limb-brightening, which would thus mimic four lobes (e.g., {\it Avery et
al.}, 1990); a single outflow but with a strong precession of the
ejection direction (e.g., {\it Ladd and Hodapp}, 1997). The angular resolution
provided by recent interferometric observations have clearly favored
the first hypothesis in at least two objects (HH\,288, {\it Gueth et al.},
2001; L\,723, {\it Lee et al.}, 2002). In both cases, the two outflows are
driven by two  independent, nearby protostars, located in the same
molecular core. It is however unclear whether the sources are
gravitationally  bound or not.

{\em 2.1.3. Time evolution}.  
There is increasing evidence that outflow collimation and morphology
changes with time 
(e.g., {\it Lee et al.}, 2002; {\it Arce and Sargent}, in preparation). 
The youngest outflows are  
highly collimated or include a
very collimated  component, strongly suggesting that jet bow
shock-driven models are appropriate to explain these objects.
Older sources present much lower collimation  factors, or --- a
somewhat more relevant parameter --- wider opening angles, pointing
towards wide-angle, wind-driven outflows (see Sec.~3.1.1). 
In fact,    
neither the jet-driven nor the wind-driven models can explain the   
range of morphological and kinematic properties that are observed   
in all outflows (see Sec.~3.2). This was noted by {\it Cabrit et al.} (1997),    
who compared outflow observations to morphologies and PV diagrams     
predicted by various hydrodynamical models. More recently, a similar  
conclusion was obtained by {\it Lee et al.} (2000, 2001, 2002) from                   
interferometric observations of 10 outflows.                       
One  attractive
scenario to reconcile all observations is to invoke the superposition
of both a jet and a wind component in the underlying protostellar wind and 
a variation in time of the relative weight between these two 
components.
One possible explanation for this scenario is that at very early ages 
only the dense collimated part of the wind can break out
of the surrounding dense infalling envelope. As the envelope
loses mass, through infall and outflow entrainment along the axis
(see Sec 4.1), the less dense and wider wind component will break through, entraining the
gas unaffected by the collimated component, and will eventually become
 the main component
responsible for the observed molecular outflow.


\bigskip
\noindent
\textbf{2.2 Outflows from high-mass protostars}
\bigskip

Outflows from more luminous protostars have received increasing
attention in recent years with the result that we now have a more
consistent understanding of massive outflow properties and their
relationship to outflows from lower luminosity objects (see, e.g.,
recent reviews by {\it Churchwell}, 1999; 
{\it Shepherd}, 2003, 2005; and {\it Cesaroni}, 2005).

Outflows from mid- to early-B type stars have mass outflow rates
$10^{-5}$ to a few $\times$ $10^{-3}$~\Msunyr, momentum rates
$10^{-4}$ to $10^{-2}$~\Msunkmsyr, and mechanical luminosity of
$10^{-1}$ to $10^2$~\Lsun.  O stars with bolometric luminosity
($L_{bol}$) of more than $10^4$~\Lsun\ generate powerful winds with
wind opening angle of about $90$\deg\ within 50~AU of the star
(measured from water masers in and along the flow boundaries and
models derived from ionized gas emission observed with resolutions of
20-100 AU, e.g., Orion: {\it Greenhill et al.}, 1998; MWC\,349A: {\it
Tafoya et al.}, 2004).  The accompanying
molecular flows can have an opening angle of more than $90$\deg
(measured from CO outflow boundaries 1000\,AU to 0.1\,pc from the
protostar).  The flow momentum rate ($> 10^{-2}$~\Msunkmsyr) is more
than an order of magnitude higher than what can be produced by stellar
winds and the mechanical luminosity exceeds $10^2$~\Lsun\ (e.g., {\it
Churchwell}, 1999; {\it Garay and Lizano}, 1999).


Outflows from early-B and late O stars can be well-collimated
(collimation factors greater than 5) when the dynamical times scale is
less than $\sim 10^4$ years.  
For a few early B (proto)stars with
outflows that have a well-defined jet, the jet appears to have
adequate momentum to power the larger scale CO flow, although this
relation is not as well established as it is for lower luminosity
sources.  For example, IRAS 20126$+$4104 has a momentum rate in the
SiO jet of $2 \times 10^{-1} \left( \frac{2 \times 10^{-9}}{{\rm
      SiO/H}_2} \right)$\,{\Msunkmsyr} while the CO momentum rate is
$6 \times 10^{-3}$\,{\Msunkmsyr} ({\it Cesaroni et al.}, 1999; 
{\it Shepherd et al.}, 2000).  Although the calculated momentum rate in the SiO jet is
adequate to power the CO flow, the uncertainties in the assumed SiO
abundance makes this difficult to prove.  Another example is
IRAS\,18151--1208 in which the {\h} jet appears to have adequate
momentum to power the observed CO flow ({\it Beuther et al.}, 2002a; 
{\it Davis et al.}, 2004).  A counter example may be the Ceph A HW2 outflow because
the momentum rate in the HCO$^+$ outflow is 20 times larger than that
of the observed ionized jet. However, the jet could be largely neutral
or there may be an undetected wide-angle wind component ({\it G\'omez et
al.}, 1999).

{\it Wu et al.} (2004) find that the average collimation factor 
 for outflows from sources with $L_{bol} > 10^3$~L{\sun} is 2.05
compared with 2.81 for flows from lower luminosity sources.  This is
true even for sources in which the angular size of the flow is at
least five times the resolution.  Table 1 of {\it Beuther and Shepherd} 
(2005) summarizes our current understanding of massive outflows from
low-spatial resolution single-dish studies and gives a summary of and
references to 15 massive flows that have been observed at higher
spatial resolution using an interferometer.  Here, we discuss a few of
these sources that illustrate specific characteristics of massive
outflows.

{\em 2.2.1. Collimated flows}. The youngest early-B protostars ($\sim 10^4$
years or less) can be jet-dominated and can have either
well-collimated or poorly collimated molecular flows.  In a few
sources, jets tend to have opening angles, $\alpha$, between 25{\deg}
and 30{\deg} but they do not re-collimate (e.g., IRAS\,20126$+$4104:
{\it Cesaroni et al.}, 1999; {\it Moscadelli et al.}, 2005; 
or IRAS\,16547$-$4247: {\it Rodr\'{\i}guez et al.}, 2005a). 
Other sources appear to generate
well-collimated jets ($\alpha \sim$ few degrees) that look like scaled
up jets from low-luminosity protostars (e.g., IRAS\,05358$+$3543:
{\it Beuther et al.}, 2002b).  All these sources are $\ltsim 10^4$ years
old --- they have not yet reached the main sequence.  In at least
one case jet activity has continued as long as $10^6$ years, although
 the associated molecular flow has a large opening angle and
complex morphology (HH80--81: {\it Yamashita et al.}, 1989; {\it Mart\'{\i} et
al.}, 1993).

One possible collimated outflow event may have been traced to a young
O5 (proto)star in the G5.89--0.39 UC HII region.  The O5 star has a
small excess at 3.5$\mu$m and is along the axis of two {\h} knots that
appear to trace a N-S molecular flow along the direction of the UC HII
region expansion ({\it Puga et al.}, 2005).  The N-S molecular flow is
unresolved so it is not clear that it is collimated even if the {\h}
knots appear to trace a collimated outflow event.  Although still
circumstantial, the evidence is mounting that the O5 star in G5.89
produced the N-S outflow and thus is forming via accretion ({\it Shepherd},
2005, and references therein).

{\em 2.2.2. Poorly collimated flows}. Poorly collimated molecular flows can
be due to: 1) extreme precession of the jet as in IRAS\,20126$+$4124
({\it Shepherd et al.}, 2000); 2) a wide-angle wind associated with a jet as
in HH\,80--81 ({\it Yamashita et al.}, 1989) or perhaps Ceph A HW2
(e.g., {\it G\'omez et al.}, 1999; {\it Rodr\'{\i}guez et al.}, 2001); 3) a strong
wide-angle wind that has no accompanying jet; or 4) an explosive event
as seen in Orion ({\it McCaughrean and Mac Low}, 1997).  In massive flows,
collimation factors as high as 4 or 5 in the molecular gas can still
be consistent with being produced by wind-blown bubbles if the cloud
core is very dense and it is easier for the flow to break out of the
cloud rather than widen the flow cavity.  Once the flow has escaped
the cloud core, the bulk of the momentum is transfered to the
inter-clump medium.

In at least some young early-B stars, both the ionized wind near the
central source and the larger scale molecular flow are poorly
collimated and there is no evidence for a well-collimated jet.
Examples of sources that do not appear to have a collimated jet
powering the flow include G192.16--3.82 ({\it Shepherd and Kurtz}, 1999, and
references therein), W75\,N VLA\,2 ({\it Torrelles et al.}, 2003, and
references therein), AFGL\,490 ({\it Schreyer et al.}, 2006, 
and references therein) 
and the SiO flow in G5.89--0.39 (not related to the O5 star
discussed above; {\it Sollins et al.}, 2004; {\it Puga et al.}, 2005).  Sources
with poorly collimated flows, no evidence for a jet and a good
determination of the dynamical age show that the ages tend to be a few
$\times 10^5$ years old and a UC HII region exists around a new ZAMS
star.

To date, extremely collimated molecular outflows have not been
observed toward sources earlier than B0.  It is possible that this is
simply a selection effect because O stars form in dense clusters and
reach the ZAMS in only a few $\times 10^4$ years.  Thus, any
collimated outflows may be confused by other flows.  In a few cases,
outflows appear to be due to a sudden explosive event such as that
seen in Spitzer images of shocked gas in G34.26$+$0.15 ({\it Churchwell},
personal communication) or the {\h} fingers of Orion.  There is now
good evidence that Source I in Orion and the Becklin-Neugebauer (BN)
object were within a few hundred AU from each other about 500 years
ago ({\it Rodr\'{\i}guez et al.}, 2005b).  Such close
encounters could disrupt the accretion process and create an explosive
outflow as seen in Orion (e.g., {\it Bonnell et al.}, 2003).

{\em 2.2.3. Evolution}. Early-B stars ($L_{bol} \sim 10^4$~\Lsun) 
generate UC HII regions and reach the ZAMS in $5-9 \times 10^4$ years
while still accreting and generating strong molecular outflows
(e.g., {\it Churchwell}, 1999; {\it Garay and Lizano}, 1999, 
and references therein).
The duration of the accretion phase is about the same as in
low-luminosity sources (e.g., $5-10 \times 10^5$ years) yet the
development of an HII region that expands to encompass the accretion
disk mid-way through the formation process suggests that there is a
sharp transition in the physical conditions at the base of the flow
where material is lifted off the surface of the disk and collimated.

Well-collimated molecular flows from massive protostars tend to be in
systems with ages less than a few times $10^4$ years old where the
central object has not yet reached the main sequence
(e.g., IRAS\,05358$+$3543 is well-collimated over approximately 1\,pc).
In these young sources the effects of increased irradiation on the disk and disk-wind
due to the stellar radiation field are minimal.  Poorly
collimated flows (opening angle greater than 50{\deg} that show no
evidence for a more collimated component) are associated with more
evolved sources that have detectable UC HII regions and the central
star has reached the main sequence.

To account for the differences seen in flow morphologies from early B
to late O stars {\it Beuther and Shepherd} (2005) proposed two possible
evolutionary sequences which could result in similar observable
outflow signatures.  In Fig.~1 we show a schematic of the proposed
sequences and explain how the observed outflow morphologies can be related to
O and B star evolution.

\begin{figure}[htb] 
\epsscale{0.9}
\plotone{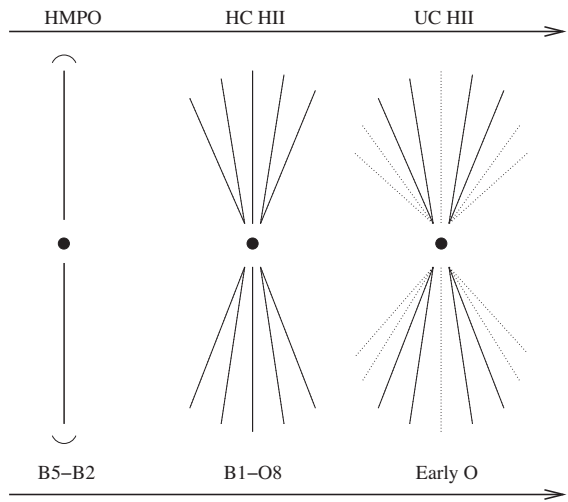}
\caption{\footnotesize Sketch of the proposed evolutionary outflow
scenario put forth by {\it Beuther and Shepherd}  (2005).  The three outflow
morphologies can be caused by two evolutionary sequences: (top) the
evolution of a typical B1-type star from a high-mass protostellar
object (HMPO) via a hyper-compact H{\sc ii} (HC\,HII) region to an
ultra-compact H{\sc ii} (UC\,HII) region, and (bottom) the evolution
of an O5-type star which goes through B1- and O8-type stages (only
approximate labels) before reaching its final mass and stellar
luminosity.  This evolutionary sequence appears to qualitatively fit
the observations, yet it must be tested against both theory and
observations.}
\end{figure}

Once a massive OB star reaches the main sequence, the increased
radiation from the central star generates significant Lyman continuum
photons and will likely ionize the outflowing gas even at large radii.
Inherently lower collimation of the ionized wind due to increased
radiation pressure is suggested by the hydrodynamic simulations of
{\it Yorke and Sonnhalter} (2002).  However the radiation pressure is still too
low by a factor of 10 to 100 to produce significant changes in the
collimation of the observed molecular flows ({\it Richer et al.}, 2000).  

The larger photon flux will also increase the ionization degree in the
molecular gas and produce shorter ion-neutral collisional timescales.
Thus, in principle, this could improve the matter-field coupling, even
aiding MHD collimation.  However, other effects are likely to
counteract this.  In particular, if the plasma pressure exceeds the
magnetic field pressure and ions are well-coupled to the field, then
the outflowing, ionized gas may be able to drag the magnetic field
lines into a less collimated configuration (see, e.g., {\it K\"onigl}, 1999;
{\it Shepherd et al.}, 2003).

Turbulence could also contribute to the decollimation of molecular
outflows from massive OB protostars.  Increased turbulence in the disk
and outflow is expected to weaken the conditions for ideal MHD and
hence weaken the collimation effect. Turbulence could be due to higher
accretion disk to stellar mass ratios ($M_{disk} > 0.3 M_\star$)
making disks susceptible to local gravitational instabilities,
increased radiation pressure and high plasma temperatures.  If the
ions and neutrals are not well coupled in a turbulent flow then ideal
MHD begins to break down and magnetic diffusivity could significantly
decollimate the molecular outflow (see, e.g., {\it Fendt and Cemeljic}, 2002).
Further, simulations by {\it Fendt and Cemeljic} find that the toroidal
magnetic field component, $B_{\phi}$, decreases with increased
turbulence. Since $B_{\phi}$ is the collimating magnetic component
(e.g., {\it Pudritz \& Banerjee}, 2005), such a decrease in $B_{\phi}$ may
contribute to the lower observed collimation for more evolved massive
molecular outflows.

\section{\textbf{MOLECULAR OUTFLOW MODELS}}

\noindent
\textbf{3.1 General Overview of Models}
\bigskip

Several outflow models have been proposed to explain how molecular
outflows from protostars are formed.  Currently, outflow models can be
separated into four broad classes ({\it Cabrit et al.}, 1997): (1)
wind-driven shells, (2) jet-driven bow shocks, (3) jet-driven
turbulent flows, and (4) circulation flows.  In the first three,
molecular outflows represent ambient material that has been entrained by
a  wide-angle wind or accelerated by a highly collimated jet.  In
the last class of models, molecular outflows are produced  by
deflected infalling gas.  Most of the work has  concentrated on
simulating outflows specifically from low-mass protostars, and little
work has been done on modeling outflows from high-mass stars.  Many
flow properties, in particular the CO spatial and velocity structure,
are broadly similar across the entire luminosity range ({\it Richer et
al.}, 2000), suggesting that similar mechanisms may be responsible for
the production of molecular outflows from both low- and high-mass
systems.  Recent results from simulation work on the disk/outflow
connection ({\it Pudritz and Banarjee}, 2005) as well as from observations
({\it Zhang et al.}, 2002; {\it Beuther et al.}, 2004) further indicate that
molecular outflows from massive stars may be approximately modeled as
 scaled-up versions of their lower mass brethren.

In the past, most
studies used analytical models to try to explain the 
outflow morphology and kinematics. However, in the last decade
computational power has increased sufficiently to 
allow for multidimensional hydrodynamical (HD) simulations of protostellar outflows
that include a simple molecular chemical network.
Numerical modeling of the molecular cooling and 
chemistry, as well as the hydrodynamics, is required in these systems,
which are described by a set of hyperbolic differential equations
with solutions that are usually mathematically chaotic and cannot be 
treated analytically.   Treatment of the
molecular cooling and chemistry facilitates a comparison of
the underlying flow with observational quantities 
(for example, the velocity distribution of mass vs.~CO intensity,
the temperature distribution of the outflowing gas, 
and the H$_2$ 1-0 S(1) maps).


{\em 3.1.1. Wind-driven shell models}. 
In the wind-driven shell model, a
wide-angle radial wind blows into the stratified surrounding ambient
material, forming a thin swept-up shell that can be identified as the
outflow shell ({\it Shu et al.}, 1991; {\it Li and Shu}, 1996; 
{\it Matzner and McKee}, 1999).
In these models, the ambient material is often assumed to be toroidal with
density $\rho_a = \rho_{ao} \sin^2\theta/r^2$, while the wind is
intrinsically stratified with density $\rho_w = \rho_{wo}/( r^2
\sin^2\theta)$, where $\rho_{ao}$ is the ambient density at the
equator and $\rho_{wo}$ is the wind density at the pole ({\it Lee et
al.}, 2001).  This class of models is attractive as it particularly
explains old outflows of large lateral extents and low collimation.

In recent years, there have been a few efforts to model
wide angle winds numerically.
{\it Lee et al.} (2001)
performed numerical HD simulations of an atomic axisymmetric
wind and compared it to 
simulations of bow shock-driven outflows.
Their wide-wind models yielded smaller values 
of $\gamma$  (see Sec.~2) over a 
narrower range (1.3--1.8), as compared to the jet models (1.5--3.5).
{\it Raga et al.} (2004b) have included both wide angle winds and bow shock models
in a study aimed at reproducing features of the southwest
lobe of HH 46/47, with the result that a jet model
is able to match enough features that they feel that 
it is not necessary to invoke a wide angle wind (although
it produces a reasonable fit to the observations).
In simulations by {\it Delamarter et al.} (2000) 
the wind is assumed to be spherical, 
even though the physical origin of such a wind is not yet clear,
and it is focused towards the polar axis 
by the density gradients in the surrounding (infalling) torus-like environment.
 In these models the low-velocity $\gamma$ ranges from approximately 1.3 to 1.5, 
 similar to other studies. The MHD simulations performed
 by {\it Gardiner et al.} (2003) show that winds that have a wide opening angle at the base
can produce a dense jet-like structure
downstream due to MHD collimation.
Very recently, axisymmetric winds have been modeled with a 
code that includes molecular chemistry and cooling as well as Adaptive 
Mesh Refinement (AMR) ({\it Cunningham et al.}, 2005). 
These last two studies produce satisfactory general outflow lobe appearance,    
however, no mass-velocity,  
position-velocity maps, or channel maps have been generated to compare with observations.

{\em 3.1.2. Turbulent jet model}. 
In the jet-driven turbulent model, 
Kelvin-Helmholtz instabilities
along the jet/environmental boundary lead to the formation of a turbulent
viscous mixing layer, through which the cloud molecular gas is entrained
({\it Cant\'o and Raga}, 1991; 
{\it Raga et al.}, 1993; {\it Stahler}, 1994;
{\it Lizano and Giovanardi}, 1995; {\it Cant\'o et al.}, 2003,
 and references therein).
The mixing layer grows both into the environment and into the jet, and
eventually the whole flow becomes turbulent. Discussion
of the few  existing numerical studies  that investigate how  
molecular outflows are created by a turbulent jet
is presented in a recent
review by {\it Raga et al.} (2004a), 
who cite the ``Torino group" as the only simulations with predictions
 for atomic (e.g., H$\alpha$, [SII]) emission 
({\it Micono et al.}, 1998). 
 The radiatively cooled jet simulations reproduce the
broken power law behavior of the observationally determined 
mass-velocity distribution, even though molecular chemistry
or cooling is not included ({\it Micono et al.}, 2000). 
However, these models produce decreasing  
molecular outflow momentum and velocity with distance from the powering source 
---opposite to that observed in most molecular outflows.
An analytical model using Kelvin-Helmhotz instabilities has recently been proposed by
{\it Watson et al.} (2004) to explain entrainment of cloud material by outflows from high-mass stars.

{\em 3.1.3. Jet  bow shock model}.
In the jet-driven bow shock model, a highly collimated jet
propagates into the surrounding ambient material, 
producing a thin outflow shell around
the jet ({\it Raga and Cabrit}, 1993; {\it Masson and Chernin}, 1993).
The physical origin of the jet is currently unclear 
and could even be considered as an extreme case of a highly  collimated
wide-angle wind without a tenuous wide-angle component.
As the jet impacts the ambient material, a pair of shocks,
a jet shock and a bow shock, are formed at the head of the jet. 
High pressure gas between the shocks is ejected sideways out of 
the jet beam, which then interacts with
unperturbed ambient gas through a broader bow shock surface,
producing an outflow shell surrounding the jet.
An episodic variation in the mass-loss rate produces
a chain of knotty shocks and bow shocks along the jet axis 
within the outflow shell. Recent analytical models without magnetic field
include {\it Wilkin} (1996), {\it Zhang and Zheng} (1997), {\it Smith et al.} (1997), 
{\it Ostriker et al.} (2001), and {\it Downes and Cabrit} (2003).

There have been two recent sets of efforts (by two different groups)
to model molecular
protostellar jets numerically in two or three spatial dimensions,
where the mass-velocity and position-velocity have routinely been
measured.  In these simulations, a tracer associated with molecular
hydrogen is followed.  However, each group approaches this problem in
a different way, with each approach having its own advantages and
disadvantages.  In an effort to resolve the post-shock region,
{\it Downes and Ray} (1999), and {\it Downes and Cabrit} (2003) have simulated
relatively low density, axisymmetric (two-dimensional) fast jets.
Alternatively, recognizing that observed flows associated with Class 0
sources have a higher density and a complex appearance, {\it Smith and
Rosen} have extended the work of {\it Suttner et al.} (1997) and {\it V\"olker et
al.} (1999) by further investigating sets of fully three-dimensional flows
(e.g., {\it Rosen and Smith}, 2004a).   The main disadvantage of this approach
is that with such high densities the post-shock region
will necessarily be under-resolved, especially in
three-dimensional flows.  Both the {\it Downes} and {\it Smith} groups
have included molecular hydrogen dissociation and reformation as well
as ro-vibrational cooling in their hydrodynamical simulations,
although the treatment of this cooling is quite different in each
group.  One example is that the {\it Downes} group turns off all cooling and
chemistry below 1000 K, while the {\it Smith and Rosen} simulations (explained in
detail in {\it Smith and Rosen}, 2003) include cooling and chemistry calculations
at essentially all temperatures (albeit with an equilibrium assumption
for some reactions). The jet flows themselves enter the grid from a limited
number of zones at one side of the computational domain, with densities and
temperatures that are constant radially (a top hat profile) and over time.
Both groups usually model the jet as nearly completely molecular
 ---even though
there are arguments suggesting that the jet will not initially be molecular,
 and that  H$_2$ might subsequently form on the internal working surfaces
of the jet ({\it Raga et al.}, 2005). The initial jet velocities of 
the {\it Downes}, and {\it Smith and Rosen} 
groups are varied with shear, pulsation, and, in the three
dimensional simulations, with precession.

These different approaches have yielded different slopes for the
computed CO intensity-velocity plots.  The {\it Downes} group results
have tended to be steeper and closer to the nominal value of
$\gamma$ = 2, while the standard {\it Rosen and Smith} case 
 has a value near 1.  Much of this difference can
be attributed to the difference in jet-to-ambient density ratio (see
 {\it Rosen and Smith}, 2004a),
which is 1 in the {\it Downes} standard case, and 10 in the 
{\it Rosen and Smith} standard case.  
The value of $\gamma$ has been shown in these simulations to evolve over
time, with steeper slopes associated with older flows.
Most of these simulations are quite young, but there has been
a recent effort to run the simulations out to t = 2300 yr
({\it Keegan and Downes}, 2005).  They
confirm the steepening of the mass-velocity slope up
to t = 1600 yr (when $\gamma$ = 1.6), and then it becomes roughly constant.
The {\it Smith and Rosen} group have investigated whether fast 
({\it Rosen and Smith}, 2004b)
or slow ({\it Smith and Rosen}, 2005) precession has an effect on
the mass-velocity slopes.  While the simulations with fast precessing
jets show a dependence of $\gamma$ on the precession angle (generally
increasing $\gamma$ with the angle), some of this dependence was 
reduced in the slowly precessing cases.  However, at this
time only very young (t $<$ 500 yr) precessing sources have been 
simulated.

The initially molecular jet simulations that include periodic velocity pulses 
exhibit position-velocity plots with a sequence of Hubble wedges, 
similar to that observed in molecular outflows produced by an episodic protostellar
wind (see Sec.~2.1.2).
Where computed, velocity channel maps in CO from molecular
jet simulations, as in {\it Rosen and Smith} (2004a), 
have a morphology similar to that of many
sources (e.g.,  HH 211,  {\it Gueth and Guilloteau}, 1999),
 i.e. revealing the knots within the jet at high 
velocities and showing the overall shape of the bow shock at low
velocities.

Some recent studies show the need to expand the interpretation 
of molecular outflow observations beyond the simulated H$_2$ and CO emission 
from the numerical models discussed above.
For example, the work of {\it Lesaffre et al.} (2004) includes 
more complex chemistry in one dimension, 
focusing on the unstable nature of combined C and J shocks.
Also, radiation transfer with a complex chemistry 
has been simulated for a steady three dimensional
(jet) flow, with a focus on HCO$^{+}$ emission ({\it Rawlings et al.}, 2004).

In addition, magnetic field effects have been included in atomic protostellar 
jets that are axisymmetric ({\it Gardiner et al.}, 2000; 
{\it Stone and Hardee}, 2000) and 
fully three-dimensional ({\it Cerqueira and de Gouveia dal Pino}, 1999, 2001)  and 
even molecular axisymmetric protostellar jets ({\it O'Sullivan and Ray}, 2000).
These studies show significant differences compared to simulations of jets
 without magnetic fields.   For example, magnetic tension, either 
along the jet axis or as a hoop stress from a toroidal field, can help collimate
and stabilize the jet ---though some of the additional stability is mitigated 
in a pulsed jet.  Some of the differences between pure HD and MHD 
simulations that show up in the axisymmetric cases are less prominent in 
three dimensional simulations ({\it Cerqueira and de Gouveia dal Pino}, 2001).

{\em 3.1.4. Circulation models}.  In circulation models the molecular
outflow is not entrained by an underlying wind or jet, it is rather
formed by infalling matter that is deflected away from the protostar
in a central torus of high MHD pressure through a quadrupolar
circulation pattern around the protostar, and accelerated above escape
speeds by local heating ({\it Fiege and Henriksen}, 1996a,b). The molecular
outflow may still be affected by entrainment from the wind or jet, but
this would be limited to the polar regions and it would not be the
dominant factor for its acceleration ({\it Lery et al.}, 1999, 2002).
Circulation models may provide a means of injecting added mass into
outflows from O stars where it appears unlikely that direct
entrainment can supply all the observed mass in the flow 
({\it Churchwell}, 1999).

The most recent numerical studies of the circulation model have focused on
a steady-state axisymmetric case, usually involving radiative heating,
magnetic fields and
Poynting flux ({\it Lery}, 2003).  
The addition of the Poynting flux in recent versions of this
model has alleviated one of its major flaws ({\it Lery et al.}, 2002), i.e. the 
inability in earlier models to generate an outflow of sufficient speed.   The
toroidal magnetic field in what is currently being called the ``steady-state
transit model'' assists in the formation of a collimated fast moving flow 
({\it Combet et al.}, 2006).

\bigskip
\noindent
\textbf{ 3.2 Comparing Observations and Models}
\bigskip

In the past ten years, molecular outflows have been mapped at high
angular resolutions with millimeter interferometers, allowing us to
confront the outflow models in great detail.   A schematic of the
predicted properties of molecular outflows produced by the different
models discussed above is presented in Fig.~2.  High-resolution
molecular outflow
observations can be used to compare the data with the outflow
characteristics shown in Fig.~2 in order to establish what model
best fits the observed outflow. 


\begin{figure}[htb]
 \epsscale{1.0}
\plotone{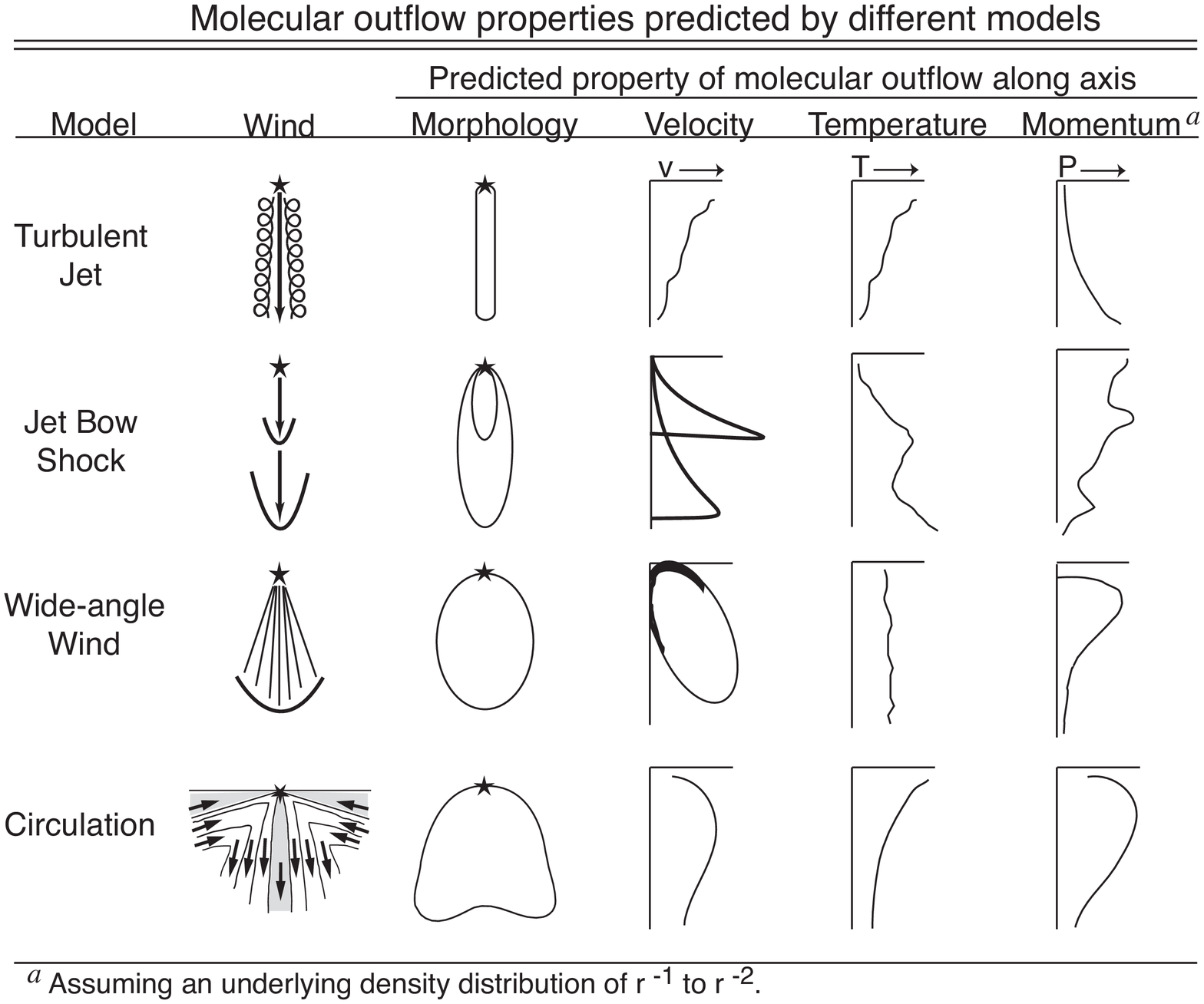}
\caption{\footnotesize Observable molecular outflow properties predicted by the four leading
broad classes of models:  1) turbulent jet 
({\it Cant\'o and Raga}, 1991; {\it Chernin and Masson}, 1995; {\it Bence et al.}, 
1996); 2) jet bow shock 
 ({\it Chernin and Masson}, 1995; {\it Cliffe et al.},1996; 
{\it Hatchell et al.}, 1999; {\it Lee et al.}, 2001);
 3) wide-angle wind ({\it Li and Shu}, 1996; {\it Lee et al.}, 2001); 
and 4) circulation models
 ({\it Fiege and Henriksen}, 1996b; {\it Lery et al.}, 1999).
In the jet-driven bow shock model, 
an episodic variation in jet velocity
produces an internal bow shock driving an internal shell, in addition to the
terminal shock.  This episodic variation can also be present in the
other wind models,  but in this figure the effects of an episodic wind
are only shown for the jet bow shock model. This figure is based on Figure 1 of 
{\it Arce and Goodman} (2002b).}
\end{figure}

Here we focus our attention on comparing observations with the
jet-driven bow shock and wide-angle wind-driven models, as most of the
numerical simulations concentrate on these two models and they are the
most promising models thus far.  The predicted mass-velocity
relationships in jet bow shock and wide-angle wind models have a slope
($\gamma$) of 1--4, in tune with observations.  Each model predicts a
somewhat different position-velocity (PV) relation that can be used to
differentiate between these two leading molecular outflow driving
mechanisms ({\it Cabrit et al.}, 1997; {\it Lee et al.}, 2000, 2001).


{\em 3.2.1. Jet-driven bow shock models vs.~observations}. Current jet-driven
bow shock  models can qualitatively account for the PV spur
structure (where the outflow velocity increases rapidly toward the
position of the internal and leading bow shocks, see Fig.~2), the
broad range of CO velocities near H$_2$ shocks, and the morphological
relation between the CO and H$_2$ emission seen in young and
collimated outflows.  These models are able to produce the observed
outflow width for highly collimated outflows, such as L\,1448, HH\,211
and HH\,212 ({\it Bachiller et al.}, 1995; 
{\it Gueth and Guilloteau}, 1999; {\it Lee et
al.}, 2001).  However, jet-driven bow shock models have difficulty
producing the observed width of poorly collimated outflows, like RNO\,91,
VLA\,05487, and L\,1221 ({\it Lee et al.}, 2000, 2001, 2002).  Jet models
produce narrow molecular outflows mainly because the shocked gas in
the bow shock working surfaces limits the transverse momentum
 (perpendicular to the jet-axis)
that can be delivered to the ambient medium.  In numerical simulations
of jets, the width of the outflow shell is mainly determined by the
effects of the leading bow shock from the jet's first impact into the
ambient material (e.g., {\it Suttner et al.}, 1997; {\it Downes and Ray}, 1999;
{\it Lee et al.}, 2001).  While the jet penetration into the cloud increases
roughly linearly with time, the width only grows as the one-third
power of time ({\it Masson and Chernin}, 1993; {\it Wilkin}, 1996;
{\it Ostriker et al.}, 2001).

Jets also have difficulty producing the observed
outflow momenta.
The transverse momentum
of the outflow shell is acquired primarily near the jet head
where the pressure gradient is large, and the  mean
transverse velocity of the shell, $\bar{v}_R$,  can be approximated by
$\bar{v}_R \simeq \beta c_s (R_j^2/R^2)$, where $R$ and $R_j$
are the outflow and jet radius, respectively, and $\beta c_s$ is the 
velocity of the gas ejected from the working surface
({\it Ostriker et al.}, 2001).
For example, in a 10,000 AU-wide molecular outflow driven by 
a 150 AU jet, and assuming $\beta c_s = 32$~km s$^{-1}$, the expected  mean
transverse velocity of the shell is only 0.03~km s$^{-1}$.
As a result, if outflows were driven by a steady jet,
the wide portions of outflow shells would exhibit extremely low velocities
and very small momenta.
This is inconsistent with the observations, especially in the wider
flows where the well-defined cavity walls have appreciable velocities
(e.g., B5-IRS1: {\it Velusamy and Langer}, 1998; 
 RNO91: {\it Lee et al.}, 2002; L1228: {\it Arce and Sargent}, 2004).
                                                                                    
Systematic wandering of the jet flow axis has been argued to
occur in several outflows
based on outflow morphology, e.g., IRAS 20126+4104 ({\it Shepherd et al.}, 2000)
and L 1157 ({\it Bachiller et al.}, 2001). 
This may mitigate the above discrepancies.
The width and momentum of the outflow shell can increase
because a wandering jet has a larger ``effective radius'' of interaction
and can impact the outflow shell more directly
({\it Raga et al.}, 1993; {\it Cliffe et al.}, 1996). Some simulations show
hints of widening by jet wandering 
({\it V\"olker et al.}, 1999; 
{\it Rosen and Smith}, 2004a; {\it Smith and Rosen}, 2005), 
but some show that a wandering jet could produce a smaller 
width than a steady jet ({\it Raga et al.}, 2004b).                                                                                   
Further calculations are needed to ascertain if motion of the jet axis
at realistic levels
can improve quantitative agreement with observed outflow features.

{\em 3.2.2.  Wide-angle wind models vs.~observations}.
Wide-angle winds can readily produce
CO outflows with large widths but have trouble producing
other commonly observed features.  
In this model, the outflow
velocity also increases with the distance from the source,
showing a lobe PV structure tilted with inclination that exhibits 
only a small velocity range at the tip.
If the tip is not observed, the PV structure appears as a 
tilted parabola (see Fig.~2).
As discussed in Sec. 3.1.1, 
most wind-driven models, assume the protostellar wind density
depends on the angle from the pole ($\theta$).
If the wind velocity has a small, or no, dependence
on $\theta$, and assuming a density stratification similar to
that proposed by {\it Li and Shu} (1996), then
the outflow width, $W$, 
can be expressed in terms of the ratio of wind to ambient  
density at the equator,
$(\rho_{wo}/\rho_{ao})$,
the wind velocity at the pole, $v_{wo}$, and the outflow age, $t$, 
as 
$W \approx (\rho_{wo}/\rho_{ao})^{1/4} v_{wo} t$
({\it Lee et al.}, 2001).
For $(\rho_{wo}/ \rho_{ao})$ between $10^{-3}$ and 10$^{-4}$,
a 100~km~s$^{-1}$ wind can produce an outflow width of 0.1 to 0.2 pc 
in 10$^4$ years. 
Thus, the wind-driven model can
produce  widths consistent with  observed molecular outflows
in about 10$^4$
years. However, these models  have problems producing
discrete bow shock type features in the entrained molecular gas, as seen in
many high-resolution maps of CO outflows (e.g., {\it Lee et al.}, 2000, 2002), 
and discrete position-velocity spur structures (and 
Hubble wedges). These features are hard to generate 
as the wide wind impacts all locations on the shell. 
Models of wide-angle pulsed winds 
produce a series of flat internal shocks 
within the outflow shell ({\it Lee et al.}, 2001), inconsistent with the
curved internal H$_2$ bow shocks typically observed in episodic
outflows (see Sec 2.1).

One possible solution to these problems
is to require the winds to have a collimated core
with a strong velocity gradient with respect to $\theta$.
A disk-wind driven from a 
large range of radii may have velocity strongly
decreasing toward equatorial latitudes,
because the asymptotic velocity on a given streamline in an MHD wind is
characteristic of the Keplerian speed at the streamline's footpoint
(see chapter by {\it Pudritz et al.}).
Further work is needed to study whether this sort of modification
can produce the observed outflow features.

{\em 3.2.3. A synthesis with an evolutionary scenario}. A model which
combines attributes of the jet and wide-angle wind models is arguably the
best match to  the available CO outflow data.  A two component
protostellar wind may be produced, for example, by a slow disk wind
and a fast central disk-driven jet or X-wind (arising from the
magnetosphere-disk boundary region).  The disk wind could help
collimate the X-wind into the jet component ({\it Ostriker}, 1997) and
provide a slow wide-angle component that drives the outflow width and
momentum (see chapter by {\it Shang et al.}).

Observational support for the synthesis model exist at different
wavelengths. 
There is mounting evidence from millimeter
observations that the morphology of some molecular outflows is better
explained with a ``dual-wind'' model (e.g., 
{\it Yu et al.}, 1999; {\it Arce and Goodman}, 2002a; 
{\it Arce and Sargent}, 2004). 
In the optical,
the forbidden emission line profiles of T Tauri stars show two velocity
components: a high-velocity component that is argued to arise in a jet
and a low-velocity component that might result from a disk-wind
({\it Kwan and Tademaru}, 1995; chapter by {\it Ray et al.}).
A possible scenario is that
the main driving agent producing most of the observed molecular outflow
may change over the time, as discussed in Sec.~2.
Numerical simulations of an evolving dual-wind model  will be critical
to study whether this proposed scenario
can reproduce the wide range of observed features
in molecular outflows from low- and high-mass protostars.

\section{\textbf{IMPACT OF OUTFLOWS ON SURROUNDING ENVIRONMENT}}

\noindent
\textbf{ 4.1 Physical Impact}
\bigskip

Outflows from newborn stars inject momentum and energy into the
surrounding molecular cloud at distances ranging  from a few AU
to up to tens of parsecs away from the source. 
 Historically, most studies  have
concentrated on the interaction between the outflow and the
surrounding core ($\sim 0.1$ to 0.3 pc) as these scales can easily be
observed with single-dish telescopes in the nearby ($\lesssim 1$~kpc )
star forming regions. More recently, studies using millimeter
interferometer array and single telescopes with focal-plane arrays
have been crucial in the understanding of the outflow's impact at
smaller ($<0.1$pc) and larger ($\gtrsim 1$ pc) scales, respectively.

{\em 4.1.1. Outflow-envelope interactions}. Protostellar winds 
originate within a few
AU of the star (see chapter by {\it Ray et al.}), and
so they are destined to interact with the dense circumstellar envelope
---the  primary mass reservoir of the forming star, with sizes in
the range of $10^3$ to $10^4$ AU.  In fact, 
survey studies of the circumstellar
gas within $10^4$ AU of low-mass YSOs show outflows contribute
significantly  to the observed mass-loss of the surrounding dense gas
(from about $10^{-8}$ to $10^{-4}$ \Msun yr$^{-1}$, 
depending on the protostar's age)
and indicate there is an evolution in the outflow-envelope interaction
 (e.g., {\it Fuller and Ladd}, 2002; {\it Arce and Sargent}, in preparation).  As shown
below, detailed studies of individual sources corroborate these
results.  The powerful outflows from low-mass class 0 sources are able
to modify the distribution and kinematics of the dense gas surrounding
a protostar, as evidenced in L\,1157 ({\it Gueth et al.}, 1997; 
{\it Beltr\'an et al.}, 2004b), and RNO\,43 ({\it Arce and Sargent}, 2005)
 where molecular line
maps show the circumstellar high-density gas has an elongated
structure and a velocity gradient, at scales of 4000 AU, along the
outflow axis. Similarly, in IRAM\,0491 ({\it Lee et al.}, 2005) 
and HH\,212 ({\it Wiseman et al.}, 2001) 
the dense gas traced by N$_2$H$^+$ and NH$_3$, respectively,
exhibit blue- and red-shifted protrusions extending along the blue and
red outflow lobes, evidence that there are strong outflow-envelope
interactions in these class 0 sources. These results clearly show that,
independent of the original (i.e., pre-protostellar outflow) 
underlying circumstellar matter distribution,  
young outflows entrain dense envelope gas along the outflow axis.  

Although
not as powerful as those of class 0 sources, the wide-angle outflows
typically observed in class I sources (with opening angles of $\gtrsim
90\arcdeg$) are capable of constraining the infalling envelope to a
limited volume outside the outflow lobes, as seen in the L\,1228 ({\it Arce
and Sargent}, 2004) and B5-IRS1 ({\it Velusamy and Langer}, 1998) outflows.
The L1228 outflow is currently eroding the surrounding envelope by
accelerating high-density ambient gas along the outflow-envelope
interface and has the potential to further widen the cavities, as the
outflow ram pressure is about a factor of 4 higher than the infall ram
pressure ({\it Arce and Sargent}, 2004). In RNO 91, a class II source, the
outflow exhibits an even wider opening angle of 160\arcdeg \/ that
is expanding, and decreasing the volume of the infall
region ({\it Lee and Ho}, 2005). 

Widening of
the outflow opening angle with age appears to be a general
trend in low-mass protostars and there is ample
evidence for erosion of the envelope due to outflow-envelope 
interactions ({\it Velusamy and Langer}, 1998; 
{\it Arce and Sargent}, 2004; {\it Arce}, 2004;
{\it Lee and Ho}, 2005;
{\it Arce and Sargent}, in preparation). 
Thus, it is clear that
even if the pre-protostellar outflow
circumstellar distribution of matter 
has a lower density along the polar regions (i.e., the outflow axis) 
as suggested by different models (i.e., 
{\it Hartmann et al.}, 1996; {\it Li and Shu}, 1996),
outflow-envelope interactions will have an 
impact on the subsequent circumstellar density distribution, as
they will help widen the cavity and constrain the infall region.   
It is tempting to extrapolate and suggest that as a young star
evolves further its outflow will eventually  
become wide enough to end the infall process and disperse
the circumstellar envelope altogether.

{\em 4.1.2. Outflow-core interactions}.
Strong evidence exists for the disruptive effects outflows have on
their parent core ---the dense gas within 0.1 to 0.3~pc of the young
star.  Direct evidence of outflow-core interaction comes from the
detection of velocity shifts in the core's medium and high-density gas
in the same sense, both in position and velocity, as the high-velocity
(low-density) molecular outflow traced by $^{12}$CO
(e.g., {\it Tafalla and Myers}, 1997; {\it Dobashi and Uehara}, 2001;
{\it Takakuwa et al.}, 2003; {\it Beltr\'an et al.}, 2004a). 
The high opacity of the $^{12}$CO lines hampers the ability to trace low-velocity
molecular outflows in high-density regions. Therefore, other 
molecular species like $^{13}$CO, 
CS, C$^{18}$O, NH$_3$, CH$_3$OH, and C$_3$H$_2$ are used to trace
the high-density gas perturbed by the underlying protostellar wind.
The average
velocity shifts
in the dense core gas are typically lower than the
average velocity of the molecular 
(\twco ) outflow, consistent with
a momentum-conserving outflow entrainment process. In addition to
being able to produce systematic velocity shifts in the gas, outflows
have been proposed to be a major source of the turbulence in the core
(e.g., {\it Myers et al.}, 1988; {\it Fuller and Ladd}, 2002; 
{\it Zhang et al.}, 2005).

Outflows can also reshape the structure of the star-forming core by
sweeping and clearing the surrounding dense gas and producing
density enhancements along the outflow axis.  The clearing 
process is
revealed by the presence of nebular emission resulting from the
scattering of photons, from the young star, off of cavity walls
created by the outflow (e.g., {\it Yamashita et al.}, 1989;
{\it Shepherd et al.}, 1998; {\it Yu et al.}, 1999), or depressions 
along the outflow axis
in millimeter molecular line maps of high density tracers
(e.g., {\it Moriarty-Schieven and Snell}, 1988; 
{\it White and Fridlund}, 1992; {\it Tafalla et
al.}, 1997). 
Outflow-induced density enhancements (and shock-heated dust)
in the core may be revealed by the dust continuum emission 
(e.g., {\it Gueth et al.}, 2003; {\it Beuther et al.}, 2004;
{\it Sollins et al.}, 2004).  A change
in the outflow axis direction with time, as observed in many sources
(see chapter by {\it Bally et al.})
will allow
an outflow to interact with a substantial volume of the core and be
more disruptive on the dense gas than outflows with a constant axis
(e.g., {\it Shepherd et al.}, 2000; {\it Arce and Goodman}, 2002a).  By accelerating
and moving the surrounding dense gas, outflows can gravitationally
unbind a significant amount of gas in the dense core thereby limiting
the star formation efficiency of the dense gas (see {\it Matzner and McKee},
2000).

The study of {\it Fuente et al.} (2002) shows that outflows {\it appear} to
be the dominant mechanism able to efficiently sweep out about 90\% of
the parent core by the end of the pre-main sequence phase of young
intermediate-mass (Herbig Ae/Be) stars.  In addition, outflows from
low- and high- mass protostars have kinetic energies comparable to the
gravitational binding energy of their parent core, suggesting outflows
have the potential to disperse the entire core (e.g., {\it Tafalla
and Myers}, 1997; {\it Tafalla et al.}, 1997).  
We may even be observing
the last stages of the outflow-core 
interaction in G192.16, a massive
(early B) young star, where the dense core gas is optically thin and
clumpy, and the ammonia core is gravitationally unstable ({\it Shepherd et
al.}, 2004).  However, further systematic observations of a statistical
sample of outflow-harboring cores at different ages are needed in
order to fully understand the details of the core dispersal mechanism
and conclude whether outflows  can disperse their entire parent
core.

Theoretical studies indicate that shocks from a protostellar wind
impacting on a dense clump of gas (i.e., a pre-stellar core) along the
outflow's path can trigger collapse and accelerate the infall process
in the impacted core ({\it Foster and Boss}, 1996; {\it Motoyama and Yoshida},
2003). Outflow-triggered star-formation has been suggested in only a
handful of sources where the morphology and velocity structure of the
dense gas surrounding a young protostar appears to be affected by the outflow
from a nearby YSO ({\it Girart et al.}, 2001; {\it Sandell and Knee}, 2001; 
{\it Yokogawa et al.}, 2003).

{\em 4.1.3. Outflow-cloud interactions far from the source}.
Giant outflows from young stars of all masses are common, 
and they can interact with
the cloud gas at distances greater than 1~pc from their source
({\it Reipurth et al.}, 1997; {\it Stanke et al.}, 2000).
Outflows from low-mass protostars are able to entrain 0.1 to 1~M solar
masses of cloud material, accelerate and enhance the linewidth of
the cloud gas ({\it Bence et al.}, 1996; {\it Arce and Goodman}, 2001b), and in some 
cases their kinetic energy is comparable to (or larger than) the turbulent
energy and gravitational binding energy of their parent cloud
({\it Arce}, 2003).  The effects of giant
outflows from intermediate- and high-mass YSOs on their surroundings
can be much more damaging to their surrounding environment.  Studies of
individual sources indicate that giant outflows are able to entrain
tens to hundreds of solar masses, induce parsec-scale velocity
gradients in the cloud, produce dense massive shells of swept-up gas
at large ($> 0.5$~pc) distances from the source, and even break the
cloud apart ({\it Fuente et al.}, 1998; {\it Shepherd et al.}, 2000; 
{\it Arce and Goodman}, 2002a; {\it Benedettini et al.}, 2004).
The limited number of studies in this field 
suggest that a single giant
outflow has the {\it potential} to have a
disruptive effect on their parent molecular cloud (e.g., {\it Arce}, 2003).
Clearly, additional observations of giant outflows and their clouds are needed
in order to quantify their disruptive potential. 

Most star formation appears in a clustered mode and so multiple outflows 
should be more disruptive on their cloud than a single star.
Outflows from a group of young stars interact with a substantial
volume of their parent cloud by sweeping up the gas and dust into
shells (e.g., {\it Davis et al.}, 1999; 
{\it Knee and Sandell}, 2000), and may be a
considerable, albeit not the major, source of energy for driving
the supersonic turbulent motions inside clouds 
({\it Yu et al.}, 2000; {\it Williams et al.}, 2003; 
{\it Mac Low and Klessen}, 2004).  It has also  been
suggested that past outflow events from a group of stars may leave
their imprint on the cloud in the form of numerous cavities (e.g., {\it Bally
et al.}, 1999; {\it Quillen et al.}, 2005). Very limited (observational and
theoretical) work on this topic exists, and further observations 
of star forming regions with different environments and at different
evolutionary stages are essential to understand the role of outflows
in the gaseous environs of young stellar clusters.

\bigskip
\noindent
\textbf{ 4.2 Shock chemistry}
\bigskip

The propagation of a supersonic protostellar wind through its
surrounding medium happens primarily via shock waves. The rapid
heating and compression of the region trigger different microscopic
processes ---such as molecular dissociation, endothermic reactions,
ice sublimation, and dust grain disruption--- which do not operate in
the unperturbed gas. The time scales involved in the heating and in
some of the ``shock chemistry'' processes are short (a few $10^{2}$ to
$10^{4}$ yr), so the shocked region rapidly acquires a chemical
composition distinct from that of the quiescent unperturbed medium.
Given the short shock cooling times ($\sim$ $10^{2}$ yr, {\it Kaufman
and Neufeld}, 1996), some of these high-temperature chemical processes
only operate at the initial stages, as the subsequent chemical
evolution is dominated by low temperature processes. This chemical
evolution, the gradual clearing of the outflow path, and the likely
intrinsic weakening of the main accelerating agent, all together make
the important signatures of the shock interaction (including some of
the chemical anomalies) vanish as the protostellar object
evolves. Chemical anomalies found in an outflow can therefore be
considered as an indicator of the outflow age (e.g., {\it Bachiller et
al.}, 2001).

The chemical impact of outflows are better studied in outflows around
Class 0 sources with favorable orientation in the sky (i.e., high
inclination with respect to the line of sight). With less confusion
than that found around massive outflows, the shocked regions of
low-mass, high-collimation outflows (which often adopt the form of
well-defined bows) are well separated spatially with respect to the
quiescent gas.  Detail studies of these ``simple'' regions can help
disentangle the effects of outflow shocks from other shocks in more
complex regions --- like in circumstellar disks, where one expects to
find outflow shock effects blended with those produced by shocks
triggered by the collapsing envelope (e.g., {\it Ceccarelli et al.},
2000).

Shocks in molecular gas can be of C-type or of J-type, depending on
whether the hydrodynamical variables change continuously across the
shock front (e.g., {\it Draine and McKee}, 1993).  C-shocks are
mediated by magnetic fields acting on ions that are weakly coupled
with neutrals, they are slow, have maximum temperatures of about
2000-3000 K, and are non-dissociative. J-shocks are typically faster,
and can reach much higher temperatures.  The critical velocity at
which the change between C- and J-regime is produced depends on
several parameters such as the pre-shock density ({\it Le Bourlot et
al.}, 2002) and the presence of charged grains ({\it Flower and Pineau
des For\^ets}, 2003), and it typically ranges from $\sim 20$ up to $\sim
50$~km~s$^{-1}$. J-shocks may also occur at relatively low velocities
when the transverse component of the magnetic field is small ({\it
Flower et al.}, 2003).
Recent infrared observations of several lines of H$_{2}$, CO,
H$_{2}$O, and OH, and of some crucial atomic lines, have made possible
the estimate of temperature and physical conditions in a relatively
large sample of outflows. It follows that the interpretation of the
data from most shocked regions require a combination of C- and
J-shocks (see {\it Noriega-Crespo}, 2002; {\it van Dishoeck}, 2004,
for comprehensive reviews). Such a combination of shocks can be
obtained by the overlap of multiple outflow episodes as observed in
several sources, and/or by the bow shock geometry which could generate
J-shocks at the apex of the bow together with C-shocks at the bow
flanks ({\it Nisini et al.}, 2000; {\it O'Connell et al.}, 2004,
2005).  C-shocks are particularly efficient in triggering a distinct
molecular chemistry in the region in which the molecules are preserved
and heated to $\sim$ 2000-3000 K.  Moreover, molecules can also reform
in J-shocked regions when the gas rapidly cools, or in warm layers
around the hottest regions. The main processes expected to dominate
this shock chemistry were discussed by {\it Richer et al.} (2000).

Comprehensive chemical surveys have been carried out in two
prototypical Class 0 sources (L1157: {\it Bachiller and
P\'erez-Guti\'errez}, 1997; BHR71: {\it Garay et al.}, 1998).  More
recent observations, including high-resolution molecular maps, have
been made for a sample of sources, for example: L1157 ({\it Bachiller et
al.}, 2001), NGC1333 IRAS2 ({\it J\o rgensen et al.}, 2004), NGC1333
IRAS 4 ({\it Choi et al.}, 2004), NGC2071 ({\it Garay et al.}, 2000),
Cep-A ({\it Codella et al.}, 2005). These observations have revealed
that there are important differences in molecular abundances in
different outflow regions.  Such variations in the abundances may be
linked to the time evolution of the chemistry ({\it Bachiller et al.},
2001) and may also be related to variations in the abundance of the
atomic carbon ({\it J\o rgensen et al.}, 2004).

SiO exhibits the most extreme enhancement factors (up to $\sim
10^{6}$) with respect to the quiescent unperturbed medium. Such high
enhancements are often found close to the heads (bowshocks), and along
the axes, of some highly collimated outflows (e.g., 
{\it Dutrey et al.}, 1997, and references therein; 
{\it Codella et al.}, 1999;
{\it Bachiller et al.}, 2001; {\it Garay et al.}, 2002; 
{\it J\o rgensen et al.}, 2004; {\it Palau et al.}, 2006, and references therein). 
Sputtering of atomic Si from the dust
grains is at the root of such high SiO abundances ({\it Schilke et al.}, 1997), 
a process which requires shock velocities in excess of
 $\sim 25$~km~s$^{-1}$.
Accordingly, the SiO lines usually present broad wings and,
together with CO, the SiO emission usually reaches the highest
terminal velocities among all molecular species. Moreover, recent
observations of several outflows have revealed the presence of a
narrow ($<$ 1~km~s$^{-1}$) SiO line component ({\it Lefloch et al.}, 1998; 
{\it Codella et al.}, 1999; {\it Jim\'enez-Serra et al.}, 2004). 
The presence of SiO at low velocities
is not well understood. Plausible explanations include that this is
the signature of a shock precursor component ({\it Jim\'enez-Serra et
al.}, 2004, 2005) or that SiO is indeed produced at high velocities and
subsequently slowed down in time scales of $\sim 10^{4}$ yr ({\it Codella
et al.}, 1999).

CH$_{3}$OH and H$_{2}$CO are also observed to be significantly
overabundant in several outflows, enhanced by factors of
about 100 ({\it Bachiller et al.}, 2001; {\it Garay et al.}, 2000; 
{\it Garay et al.}, 2002; 
{\it J\o rgensen et al.}, 2004; {\it Maret et al.}, 2005).
These two species are likely evaporated directly
from the icy dust mantles, and in many cases the terminal velocities of their line
profile wings are significantly lower than that of SiO, probably
because CH$_{3}$OH and H$_{2}$CO do not survive at velocities as high
as those required to form SiO ({\it Garay et al.}, 2000). Thus, 
an enhancement of CH$_{3}$OH and H$_{2}$CO with no SiO may
indicate the existence of a weak shock. 
On the other hand, after the passage of a strong 
shock, and once the abundances of CH$_{3}$OH, H$_{2}$CO and SiO are 
enhanced in the gas phase, 
one would expect the SiO molecules to re-incorporate to the grains 
while some molecules of CH$_{3}$OH and H$_{2}$CO remain in 
the gas-phase, as these two molecules
are more volatile than SiO (their molecular depletion
timescales are about a few $10^{3}$ yr for densities of $\sim 10^{6}$
cm$^{-3}$). In this scenario enhancement of CH$_{3}$OH 
and H$_{2}$CO most likely may mark a later stage in the shock evolution than that traced by 
high SiO abundances. 

In several outflows HCO$^{+}$ high velocity emission is only prominent
in regions of the outflow which are relatively close to the driving
sources ({\it Bachiller et al.}, 2001; {\it J\o rgensen et al.},
2004). In such regions, the HCO$^{+}$ abundance can be enhanced by a
factor of $\sim$20. This behavior can be understood if the HCO$^{+}$
that was originally produced through shock-induced chemistry (e.g.,
{\it Rawlings et al.}, 2004) is destroyed by dissociative
recombination or by reaction with the abundant molecules of H$_{2}$O
({\it Bergin et al.}, 1998).  Once the abundance of the gaseous
H$_{2}$O decreases due to freeze-out, the abundance of HCO$^{+}$ may
increase. A rough anti-correlation between CH$_{3}$OH and HCO$^{+}$
({\it J{\o}rgensen et al.}, 2004) seems to support these arguments.
In other cases, HCO$^{+}$ emission is observed at positions close to
HH objects that can be relatively distant from the driving sources. In
fact, together with NH$_{3}$, HCO$^{+}$ is expected to be enhanced in
clumps within the molecular cloud by UV irradiation from bright HH
objects ({\it Viti and Williams}, 1999), an effect observed near HH2
according to {\it Girart et al.}  (2002). Nevertheless, {\it Girart et
al.} (2005) have recently found that UV irradiation alone is
insufficient to explain the measured HCO$^{+}$ enhancements and that
strong heating (as that caused by a shock) is also needed.

The chemistry of sulfur bearing species is of special interest as it
has been proposed to be a potential tool to construct chemical clocks
to date outflows (and hence their protostellar driving sources). The
scenario initially proposed by a number of models is that H$_{2}$S is
the main reservoir of S in grain mantles, although recent observations
seem to indicate that OCS is more abundant on ices than H$_{2}$S ({\it
Palumbo et al.}, 1997; {\it van der Tak et al.}, 2003).
Once H$_{2}$S is ejected to the gas phase by the effect of shocks, its
abundance will rapidly decrease after $10^4$~yr (e.g., {\it Charnley},
1997) due to oxidation with O and OH, thereby producing SO (first) and
SO$_{2}$ (at a later time).  Models and observations indicate that the
SO/H$_{2}$S and SO$_{2}$/H$_{2}$S ratios are particularly promising
for obtaining the relative age of shocks in an outflow ({\it
Charnley}, 1997; {\it Hatchell et al.}, 1998; {\it Bachiller et al.},
2001; {\it Buckle and Fuller}, 2003).  On the other hand, recent
models by {\it Wakelam et al.} (2004) have shown that the chemistry of
sulfur can be more complex than previously thought since --- among
other reasons --- the abundances of the sulfur-bearing species
critically depend on the gas excitation conditions, which in turn
depend on the outflow velocity structure.  {\it Wakelam et al.} (2005)
used the SO$_{2}$/SO and the CS/SO ratios to constrain the age of the
NGC1333 IRAS2 outflow to $\leq 5$ x $10^{3}$ yr. A recent study by
{\it Codella et al.} (2005) confirms that the use of the SO/H$_{2}$S
and SO$_{2}$/H$_{2}$S ratios is subject to important uncertainties
in many circumstances, and that other molecular ratios
(e.g., CH$_{3}$OH/H$_{2}$CS, OCS/H$_{2}$CS) can be used as more
effective chemical clocks to date outflows.

Recent work has revealed that chemical studies can be useful for the
investigation of interstellar gas structure.  For instance, {\it Viti
et al.} (2004) have recently shown that, if the outflow chemistry is
dominated by UV irradiation, clumping in the surrounding medium prior
to the outflow passage is needed in order to reproduce the observed
chemical abundances in some outflows.  We stress, however, that this
result depends on the chemical modeling and that more work is needed
before it can be generalized.

\section{\textbf{FUTURE WORK}}

We discussed how the high angular resolution observations have revealed 
general properties and evolutionary trends in molecular outflows from
low- and high-mass protostars. However,  these results rely on a 
limited number of outflows maps, thus making any statistical 
analysis somewhat dangerous. A large sample of fully mapped
outflows  at different evolutionary stages, using millimeter interferometers, 
is needed to soundly establish an empirical model of outflow evolution,
and the outflow's physical and chemical impact on its surroundings. Also, 
detail mapping of many outflows
will enable a thorough 
comparison with different  numerical outflow models in order to study
the outflow entrainment process.

Further progress in our understanding of outflows is expected
from current or planned instrument developments that aim at improving
both the sensitivity and the angular resolution, while opening new
frequency windows.  The soon to be implemented improvements to the
IRAM Plateau de Bure interferometer --- which include longer baselines,
wider frequency coverage, and better sensitivity --- as well as the
soon to be operational Combined Array for Research in Millimeter-wave
Astronomy (CARMA)  will allow multi-line large-scale mosaic maps
with $1"$ resolution (or less), required to thoroughly study the
outflow physical properties (e.g., kinematics, temperature,
densities), the entrainment process and the different chemical
processes along the outflows' entire extent. In addition, large-scale
mosaic maps of clouds with outflows will allow the study of the impact
of many outflows on their parent cloud.  The Atacama Large Millimeter
Array (ALMA), presumably operational by 2012, will have the ability to
 determine high-fidelity kinematics and morphologies
of even the most distant outflows in our Galaxy as well as flows in
near-by galaxies.  The superb (sub-arcsecond) angular resolution will
be particularly useful to study how outflows are ejected from
accretion disks, how molecular gas is entrained in the outflow, and
the interaction between the molecular jet/outflow and the environment
very close to the protostar (i.e., the infalling envelope, and
protoplanetary disk).  The Expanded Very Large Array (EVLA), expected
to be complete in 2012, will be critical to image the wide-opening
angle, ionized outflow close to the powering source, and will allow
sensitive studies of re-ionization events in
 jets, {\water} masers and SiO(1--0) in outflows.

New submillimeter facilities and telescopes under construction will
soon provide sensitive observations of high excitation lines,
important for the study of outflow driving and entrainment, as well as
shock-induced chemical processes.  The recently dedicated
Submillimeter Array (SMA) is the first instrument capable of studying
the warm molecular gas in the CO(6--5) line, at (sub)arcsecond
resolution, allowing to trace the outflow components closer to the
driving source and closer to the jet axis than previously
possible. Furthermore, the large bandwidth of the correlator allows
for simultaneous multi-line observations crucial for studying the
various shock chemistry processes in the outflow.  Also, the Herschel
Space Observatory (HSO) will measure the abundances of shock tracers
of great interest, in particular water, which cannot be observed from
the ground.

In the near future, greater computing power will make possible larger
scale numerical simulations that take advantage of adaptive grids,
better and more complex cooling and chemistry functions, and the
inclusion of radiative transfer and magnetic fields.  Given the wealth
of high-resolution data that will soon be available, numerical studies
will need to compare the simulated outflows with observations in more
detail, using the outflow density, kinematics, temperature and chemical
structure.  In addition, simulations that run for far longer times
($\sim 10^4 - 10^5$~yr) than current models ($\sim 10^3$~yr) are
needed to study the outflow temporal behavior and evolution.  Advances
in computing, perhaps including GRID technology, may even allow a
version of a virtual telescope, where both numerical modelers and
observers can find the best fit from a set of models for different
sources. \\

\textbf{ Acknowledgments.} 
H.G.A.~is supported by an NSF Astronomy and Astrophysics 
Postdoctoral Fellowship under award AST-0401568.
D.S.~is supported by the National Radio Astronomy Observatory, 
a facility of the National Science Foundation 
  operated under cooperative agreement 
  by Associated Universities, Inc.
R.B.~acknowledges partial support from Spanish grant AYA2003-7584.
A.R.~acknowledges the support
of the Visitor Theory Grant at Armagh Observatory, which hosted the author
while some of the review was written.
H.B.~acknowledges financial support by the
Emmy-Noether-Program of the Deutsche Forschungsgemeinschaft (DFG,
grant BE2578).

\bigskip

\centerline\textbf{ REFERENCES}
\bigskip
\parskip=0pt
{\small
\baselineskip=11pt
\refs Anglada G., Estalella R., Rodr\'{i}guez L. F., Torrelles J. M.,
      Lopez R., and Cant\'o, J. (1991) {\em \apj, 376}, 615-617.

\refs Arce H. G. (2003)  {\em  \rmxaa \/ Conf. Series, 15}, 123-125.

\refs Arce H. G. (2004) In 
{\em IAU Symp.~221: Star Formation at High Angular Resolution} 
(M. Burton et al., eds.) pp. 345-350. Kluwer, Dordrecht.

\refs Arce H. G. and Goodman A. A. (2001a) {\em \apj, 551}, L171-L174.

\refs Arce H. G. and Goodman A. A. (2001b) {\em \apj, 554}, 132-151. 

\refs Arce H. G. and Goodman A. A. (2002a) {\em \apj, 575}, 911-927. 

\refs Arce H. G. and Goodman A. A. (2002b) {\em \apj, 575}, 928-949.

\refs Arce H. G. and Sargent A. I. (2004) {\em \apj, 612}, 342-356.

\refs Arce H. G. and Sargent A. I. (2005) {\em \apj, 624}, 232-245.


\refs Avery L. W., Hayashi S. S., and White G. L. (1990) {\em \apj,
      357}, 524-530.

      
\refs Bachiller R. and Tafalla M. (1999) In {\em The Origin of
      Stars and Planetary System} (C. J. Lada and N. D. Kylafis, eds.), 
      pp. 227-265. Kluwer, Dordrecht.

\refs Bachiller R., Mart\'{i}n-Pintado J., Tafalla M., Cernicharo J., and
Lazareff B. (1990) {\em \aap, 231}, 174-186.

 \refs Bachiller R., Guilloteau S., Dutrey A., Planesas P., and
 Mart\'{i}n-Pintado J. (1995) {\em \aap, 299}, 857-868.

\refs Bachiller R. and P\'erez-Guti\'errez M. (1997) {\em \apj, 487}, L93-L96.

\refs Bachiller R., Guilloteau S., Gueth F., Tafalla M., Dutrey
      A., Codella C., and Castets A. (1998) {\em \aap, 339},
      L49-L52.

      
\refs Bachiller R., Gueth F., Guilloteau S., Tafalla M., and 
      Dutrey A. (2000) {\em \aap, 362}, L33-L36.

\refs Bachiller R., P\'erez-Guti\'errez M., 
   Kumar M. S. N., and Tafalla, M. (2001) {\em \aap, 372}, 899-912.

\refs Bally  J. and Lada C. J. (1983) {\em \apj, 265}, 824-847.

\refs Bally J., Reipurth B., Lada C. J., and Billawala Y. (1999) {\em \aj, 117}, 410-428.
 
\refs Beltr\'an M. T., Girart J. M., Estalella R., and Ho P. T. P. (2004a) {\em \aap, 426}, 941-949.

\refs Beltr\'an M. T., Gueth F., Guilloteau S., and Dutrey A. (2004b) {\em \aap, 416}, 631-640.

\refs Bence S. J., Richer J. S., and Padman R. (1996) {\em \mnras, 279}, 866-883.

\refs Benedettini M., Molinari S., Testi L., and Noriega-Crespo A. (2004) {\em \mnras, 347}, 295-306.     

\refs Bergin E. A., Neufeld D. A., and Melnick G. J. (1998) {\em \apj, 499}, 777-792.

\refs Beuther H. and Shepherd, D. S. (2005) In {\em Cores to Clusters: Star 
Formation with Next Generation Telescopes} (M.S.N. Kumar et al., eds.), 
pp. 105-119. Springer, New York.

\refs Beuther H., Schilke P., Sridharan T. K., Menten K. M.,
      Walmsley C. M., et al. 
      (2002a) {\em \aap, 383}, 892-904.  

\refs Beuther H., Schilke P., Gueth F., McCaughrean M., Andersen M., et al.
      (2002b) {\em \aap, 387}, 931-943.

\refs Beuther H., Schilke  P., and Gueth F. (2004) {\em \apj, 608}, 330-340.


\refs Bonnell  I. A., Bate M. R., and Vine S. G. (2003) {\em \mnras, 343}, 413-418.  

\refs Bontemps S., Andr\'e P., Terebey S., and Cabrit S. (1996)  {\em \aap, 311}, 858-872.


\refs Buckle J. V. and  Fuller G. A. (2003) {\em \aap, 399}, 567-581.

\refs Cabrit S. and Bertout C. (1992) {\em \aap, 311}, 858-872.

\refs Cabrit S., Raga A., and Gueth F. (1997) In 
{\em IAU Symp.~182: Herbig-Haro Flows and the Birth 
of Stars} (B. Reipurth and C. Bertout, eds.) 
pp. 163-180. Kluwer, Dordrecht.

\refs Cant\'o J. and Raga A. C. (1991) {\em \apj, 372}, 646-658. 
 
\refs Cant{\'o} J., Raga A. C., and Riera A. (2003) {\em \rmxaa, 39}, 207-212. 

\refs Ceccarelli C., Castets A., Caux E., Hollenbach D., Loinard L., et al.
(2000) {\em \aap, 355}, 1129-1137.

\refs Cernicharo J., Neri R., and Reipurth B. (1999) In 
      {\em IAU Symp.~182: Herbig-Haro Flows and the Birth of Low Mass Stars} 
      (B. Reipurth and C. Bertout, eds.), pp. 141-152. 
      Kluwer,  Dordrecht.

\refs Cerqueira A. H. and de Gouveia dal Pino E. M. (1999) {\em \apj, 510}, 828-845.

\refs Cerqueira A. H. and de Gouveia dal Pino E. M. (2001) {\em \apj, 560}, 779-791.

\refs Cesaroni R. (2005) {\em \apss, 295}, 5-17.  

\refs Cesaroni R., Felli M., Jenness T., Neri R., Olmi L., et al.
      (1999) {\em \aap, 345}, 949-964.

\refs Chandler C. J. and Richer J. S. (2001) {\em \apj, 555}, 139-145.

\refs Charnley S. B. (1997) {\em \apj, 481}, 396-405.

\refs Chernin L. M. and Masson C. R. (1995) {\apj, 455}, 182-189.


\refs Choi M. (2005) {\em \apj, 630}, 976-986.

\refs Choi M., Kamazaki T., Tatematsu K., and  Panis J. F. (2004)
    {\em \apj, 617}, 1157-1166.
    
\refs Churchwell E. (1999) In {\em
      The Origin of Stars and Planetary Systems} (C. J. Lada and
      N. D. Kylafis, eds.) pp. 515-552. Kluwer, Dordrecht.


\refs Cliffe J. A., Frank A., and Jones T. W. (1996) {\em \mnras, 282}, 
   1114-1128.


\refs Codella C., Bachiller R., and Reipurth B. (1999) {\em \aap,
      343}, 585-598.

\refs Codella C., Bachiller R., Benedettini M., Caselli P., Viti S.,
 and Wakelam V. (2005) {\em \mnras, 361}, 244-258.
          
\refs Combet C., Lery T., and Murphy G. C. (2006) {\em \apj, 637}, 798-810. 

\refs Cunningham A., Frank A., and Hartmann L. (2005) {\em \apj, 631}, 1010-1021.

\refs Davis C. J., Matthews H. E., Ray T. P., Dent W. R. F., and
      Richer J. S. (1999) {\em \mnras, 309}, 141-152. 

\refs Davis C. J., Varricatt W. P., Todd S. P., and Ramsay Howat
      S. K. (2004)  {\em \aap, 425}, 981-995.

\refs Delamarter G., Frank A., and Hartmann L. (2000) {\em \apj, 530}, 923-938. 

\refs Dobashi K. and Uehara H. (2001) {\em \pasj, 53}, 799-809.

\refs Downes T. P. and Cabrit S. (2003) {\em \aap, 403}, 135-140.

\refs Downes T. P. and Ray T. P. (1999) {\em \aap, 345}, 977-985.


\refs Draine B. T. and McKee C. F. (1993) {\em Ann. Rev. Astron. Astrophys., 31}, 373-432.

\refs Dutrey A., Guilloteau S., and Bachiller R. (1997) {\em \aap,
      325}, 758-768.

\refs Eisl\"offel J., Smith M. D., Christopher J., and Ray T. P. (1996)
      {\em \aj, 112}, 2086-2093.


\refs Fendt C. and Cemeljic M. (2002) {\em \aap, 395}, 1045-1060.

\refs Fiege J. D. and Henriksen R. N. (1996a) {\em \mnras, 281}, 1038-1054. 
 
\refs Fiege J. D. and Henriksen R. N. (1996b) {\em \mnras, 281}, 1055-1072. 

\refs Flower D. R. and Pineau des For\^ets G. (2003) {\em \mnras, 343}, 390-400.

\refs Flower D. R., Le Bourlot J., Pineau des For\^ets G., 
and Cabrit S. (2003) {\em \apss, 287}, 183-186.
 
\refs Foster P. N. and Boss A. P. (1996) {\em \apj, 468}, 784-796.


\refs Fuente A., Mart\'{i}n-Pintado J., Rodr\'{i}guez-Franco A.,
and Moriarty-Schieven G. D. (1998) {\em \aap, 339}, 575-586. 

\refs Fuente A., Mart\'{i}n-Pintado J., Bachiller R.,
Rodr\'{i}guez-Franco A., and Palla F. (2002) {\em \aap, 387},
977-992.

\refs Fuller G. A. and Ladd E. F. (2002) {\em \apj, 573}, 699-719.

\refs Garay G., K\"ohnenkamp I., Bourke T. L., Rodr\'{i}guez L. F., 
and Lehtinen K. K. (1998)  {\em \apj, 509}, 768-784.

\refs Garay G. and Lizano S. (1999) {\em \pasp, 111}, 1049-1087.

\refs Garay G., Mardones D., and  Rodr\'{i}guez L. F. (2000)  
    {\em \apj, 545}, 861-873.

\refs Garay G., Mardones D., Rodr\'{i}guez L. F., Caselli P., and
Bourke T. L. (2002) {\em \apj, 567}, 980-998. 

\refs Gardiner T. A., Frank A., Jones T. W., and Ryu D. (2000) {\em \apj, 530}, 834-850. 

\refs Gardiner T. A., Frank A., and Hartmann L. (2003) {\em \apj, 582}, 269-276. 

\refs Gibb A. G., Richer J. S., Chandler C. J., and Davis C. J. (2004)
      {\em \apj, 603}, 198-212.


\refs Girart J.~M., Estalella R., Viti S., Williams D. A., and Ho P. T. P. (2001) {\em \apj, 562},
   L91-L94.
   
\refs Girart J. M., Viti S., Williams D. A., Estalella R., and Ho P. T. P. (2002) {\em \aap, 388}, 1004-1015.

\refs  Girart J. M., Viti S., Estalella R., and Williams D. A. (2005) {\em \aap, 439}, 601-612.

\refs G\'omez J. F., Sargent A. I., Torrelles J. M., Ho P. T. P.,
      Rodr\'{\i}guez L. F., et al.
      (1999)  {\em \apj, 514}, 287-295.
   
\refs Greenhill L. J., Gwinn C. R., Schwartz C., Moran J. M., and
      Diamond P. J. (1998) {\em Nature, 396}, 650-653. 

\refs Gueth F. and Guilloteau S. (1999) {\em \aap, 343}, 571-584.

\refs Gueth F., Guilloteau S., and Bachiller, R. (1996) {\em \aap,
      307}, 891-897.

\refs Gueth F., Guilloteau S., Dutrey A., and Bachiller R. (1997) {\em \aap, 323}, 943-952.

\refs Gueth F., Guilloteau S., and Bachiller R. (1998) {\em \aap,
      333}, 287-297.

\refs Gueth F., Schilke P., and McCaughrean M. J. (2001) {\em
      \aap, 375}, 1018-1031.

\refs Gueth F., Bachiller R., and Tafalla M. (2003) {\em \aap, 401}, L5-L8.


\refs Hartmann L., Calvet N., and Boss A. (1996) {\em \apj, 464}, 387-403.

\refs Hatchell J., Thompson M. A., Millar T. J., and MacDonald, G. H. (1998) 
         {\em \aap, 338}, 713-722.

\refs Hatchell J., Fuller G. A., and Ladd E. F. (1999) {\em \aap, 344}, 687-695.

\refs Hirano N., Liu S.-Y., Shang H., Ho T. P. T., Huang H.-C., et al. 
 (2006) {\em \apj, 636}, L141-L144. 

\refs Jim\'enez-Serra I., Mart\'{i}n-Pintado J., Rodr\'{i}guez-Franco A., and
Marcelino N.~(2004) {\em \apj, 603}, L49-L52.

\refs Jim\'enez-Serra I., Mart\'{i}n-Pintado J., Rodr\'{i}guez-Franco A., and
Mart\'{i}n S. (2005) {\em \apj, 627}, L121-L124.

\refs J\o rgensen J. K., Hogerheijde M. R., Blake G. A., van Dishoeck E. F., 
 Mundy L. G., and Sch\"oier F. L. (2004) {\em \aap, 416}, 1021-1037.

\refs Kaufman M. J. and  Neufeld D. A. (1996) {\em \apj, 456}, 611-630.
    
\refs Keegan R. and Downes T. P. (2005) {\em \aap, 437}, 517-524.

\refs Knee L. B. G. and Sandell G. (2000) {\em \aap, 361}, 671-684.

\refs K\"onigl A. (1999) {\em \nar, 43}, 67-77.
      

\refs Kwan J. and Tademaru E. (1995) {\em \apj, 454}, 382-393.

\refs Lada C. J. and Fich M. (1996) {\em \apj, 459}, 638-652.
 
\refs Ladd E. F and Hodapp K. W. (1997) {\em \apj, 474}, 749-759. 


\refs Le Bourlot J., Pineau des For\^ets G., Flower D. R., and Cabrit S. (2002) {\em \mnras, 332}, 985-993.

\refs Lee C.-F. and Ho P. T. P. (2005) {\em \apj, 624}, 841-852.

\refs Lee C.-F., Mundy L. G., Reipurth B., Ostriker E. C., and
      Stone J. M. (2000) {\em \apj, 542}, 925-945.

\refs Lee, C.-F., Stone J. M., Ostriker E. C., and Mundy L. G. (2001)
  {\em \apj, 557}, 429-442. 

\refs Lee C.-F., Mundy L. G., Stone J. M., and Ostriker E. C. (2002)
      {\em \apj, 576}, 294-312.
 
\refs Lee C.-F., Ho P. T. P., and White S. M. (2005) {\em \apj, 619}, 948-958.

\refs Lefloch B., Castets A., Cernicharo J., and  Loinard L. (1998) {\em \apj, 504}, L109-L112.


\refs Lery T. (2003)  {\em \apss, 287}, 35-38.

\refs Lery T., Henriksen R. N., and Fiege J. D. (1999) {\em \aap, 350}, 254-274.

\refs Lery T., Henriksen R. N., Fiege J. D., Ray T. P., Frank A., and 
Bacciotti F. (2002) {\em \aap, 387}, 187-200.
 
\refs Lesaffre P., Chi\`eze J.-P., Cabrit S., and Pineau des For\^ets G. (2004) {\em \aap, 427}, 147-155.


\refs Li Z.-Y. and Shu F. H. (1996)  {\em \apj, 472}, 211-224.

\refs Lizano S. and Giovanardi C. (1995) {\em \apj, 447}, 742-751.
 
\refs Mac Low M.-M. and Klessen R. (2004) {\em Rev. Modern Phys., 76}, 125-196.


\refs Maret S., Ceccarelli C., Tielens A. G. G. M., Caux E., Lefloch B., et al.
(2005) {\em \aap, 442}, 527-538.

\refs Mart\'{\i}n-Pintado J., Bachiller R., and Fuente A. (1992) {\em
      \aap, 254}, 315-326.

\refs Mart\'{\i} J., Rodr\'{\i}guez L.~F., and Reipurth B. (1993) {\em \apj, 416}, 208-217.

\refs Masson C. R. and Chernin L. M. (1993) {\em \apj, 414}, 230-241. 

\refs Matzner C. D. and McKee C. F. (1999) {\em \apj, 526}, L109-L112. 
       
\refs Matzner C. D. and McKee C. F. (2000) {\em \apj, 545}, 364-378.

\refs McCaughrean M. J. and Mac Low, M.-M. (1997) {\em \aj, 113}, 391-400. 
      

\refs Micono M., Massaglia S., Bodo G., Rossi P., and Ferrari A. (1998) {\em \aap, 333}, 
   1001-1006.

\refs Micono M., Bodo G., Massaglia S., Rossi P., and Ferrari A. (2000) 
{\em \aap, 364}, 318-326. 

\refs Moriarty-Schieven G. H. and Snell R. L. (1988) {\em \apj, 332}, 364-378.

\refs Moscadelli L., Cesaroni R., and Rioja M. J. (2005)  {\em \aap, 438}, 889-898.

\refs Motoyama, K. and Yoshida T. (2003) {\em \mnras, 344}, 461-467.

\refs Myers P. C., Heyer M., Snell R. L., and Goldsmith P. F. (1988) {\em \apj, 324}, 907-919.


\refs Nisini B., Benedettini M., Giannini T., and Codella C. (2000)
    {\em \aap, 360}, 297-310.

\refs Noriega-Crespo A. (2002)  {\em \rmxaa \/ Conf. Ser., 13}, 71-78.
    
\refs O'Connell B., Smith M. D., Davis C. J., Hodapp K. W., Khanzadyan T., and
   Ray T. (2004) {\em \aap, 419}, 975-990. 
          
\refs O'Connell B., Smith M. D., Froebrich D., Davis C. J., and Eisl\"offel J. (2005) 
    {\em \aap, 431}, 223-234.
    
\refs Ostriker E. C. (1997)  {\em \apj, 486}, 291-306.
  
\refs Ostriker E. C., Lee C.-F., Stone J. M., and Mundy L. G. (2001)
{\em \apj, 557}, 443-450.  

\refs O'Sullivan S. and Ray T. P. (2000) {\em \aap, 363}, 355-372. 

\refs Palau A., Ho P. T. P., Zhang Q., Estalella R., Hirano N., et al.
   (2006) {\em \apj, 636}, L137-L140.

\refs Palumbo M. E., Geballe T. R., and  Tielens A. G. G. M. (1997) 
    {\em \apj, 479}, 839-844.

\refs Pudritz R. E. and Banerjee R. (2005) In {\em IAU
      Symp.~227: ``Massive Star Birth: A Crossroads of
      Astrophysics''} (R. Cesaroni et al., eds.), pp. 163-173. 
      Cambridge Univ., Cambridge. 

\refs Puga E., Feldt M., Alvarez C., and Henning T.  (2005) {\em
      Poster presented at IAU Symp. 227: 
       ``Massive Star Birth: A Crossroads of Astrophysics''}.

\refs Quillen A. C., Thorndike S. L., Cunningham A., Frank A., Gutermuth R. A., et al.
  (2005)  {\em \apj, 632}, 941-955.

\refs Raga A. C. and Cabrit S. (1993) {\em \aap, 278}, 267-278.

\refs Raga A. C., Cant\'o J., Calvet N., Rodr\'{i}guez L. F., and Torrelles 
J. M. (1993) {\em \aap, 276}, 539-548.

\refs Raga A. C., Cabrit S., and Cant\'o J. (1995) {\em \mnras, 273}, 422-430.

\refs Raga A. C., Beck T., and  Riera A. (2004a) {\em \apss, 293}, 27-36.
        
\refs Raga A. C., Noriega-Crespo A., Gonz{\'a}lez R. F., and Vel{\'a}zquez P. F. (2004b)
 {\em \apjs, 154}, 346-351. 

\refs Raga A. C., Williams D. A., and Lim A. (2005) {\em \rmxaa, 41}, 137-146.

\refs Rawlings J. M. C., Redman M. P., Keto E., and Williams D. A. (2004) {\em \mnras, 351},
1054-1062.

\refs Reipurth B., Bally J., and Devine D. (1997) {\em \aj, 114}, 2708-2735.

\refs Richer R. S., Shepherd D. S., Cabrit S., Bachiller R., and
      Churchwell E. (2000) In {\em Protostars and Planets IV}, 
      (V. Mannings et al., eds.), pp. 867-894. 
      Univ. of Arizona, Tucson.

\refs Ridge N. A. and Moore T. J. T. (2001) {\em \aap, 378}, 495-508.

\refs Rodr\'{\i}guez L. F., Carral P., Moran J. M., and Ho
      P. T. P. (1982) {\em \apj, 260}, 635-646.

\refs Rodr\'{\i}guez L. F., Torrelles J. M., Anglada G., and
      Mart\'{\i} J. (2001) {\em \rmxaa, 37}, 95-99.

\refs Rodr\'{\i}guez L. F., Garay G., Brooks K. J., and Mardones
      D. (2005a) {\em \apj, 626}, 953-958.
     
\refs Rodr\'{\i}guez L. F., Poveda A., Lizano S., and Allen C. (2005b) 
      {\em \apj, 627}, L65-L68.

\refs Rosen A. and Smith M. D. (2004a) {\em \aap, 413}, 593-607.

\refs Rosen A. and Smith M. D. (2004b) {\em \mnras, 347}, 1097-1112.  

\refs Sandell G. and Knee L. B. G. (2001) {\em \apj, 546}, L49-L52.

\refs Schilke P., Walmsley C. M., Pineau des For\^ets G., and Flower 
      D. R. (1997) {\em \aap, 321}, 293-304.
      
\refs Schreyer K., Semenov D.,  Henning T., and Forbrich J. (2006) {\em \apj, 637}, L129-L132. 

\refs Shepherd  D. S. (2003) In {\em ASP Conf. Ser. 287: ``Galactic
      Star Formation Across the Stellar Mass Spectrum''} (J. M. De Buizer 
      and N. S. van der Bliek, eds.), pp. 333-344. Astronomical
      Society of the Pacific, San Francisco. 

\refs Shepherd D. S. (2005) Massive Molecular Outflows. 
      In {\em IAU Symp.~227: ``Massive
      Star Birth: A Crossroads of Astrophysics''} (R. Cesaroni et al.,
      eds.), pp. 237-246. Cambridge Univ., Cambridge. 
      
\refs Shepherd D. S. and Kurtz S. E. (1999)  {\em \apj, 523}, 690-700.

\refs Shepherd D. S., Watson A. M., Sargent A. I., and Churchwell E. (1998) {\em \apj, 507}, 861-873.

\refs Shepherd D. S., Yu K. C., Bally J., and Testi L. (2000) {\em \apj, 535}, 833-846.

\refs Shepherd D. S., Borders T., Claussen M., Shirley Y., and Kurtz S. (2004) 
   {\em \apj, 614}, 211-220.

\refs Shu F. H., Ruden S. P., Lada C. J., and Lizano S. (1991) {\em \apj, 370}, L31-L34. 

      
\refs Smith M. D. and Rosen A. (2003) {\em \mnras, 339}, 133-147.
      
\refs Smith M. D. and Rosen A. (2005) {\em \mnras, 357}, 579-589.
 
\refs Smith M. D., Suttner G., and Yorke H. W. (1997) {\em \aap, 323}, 223-230. 

\refs Sollins P. K., Hunter T. R., Battat J., Beuther H., Ho P. T. P., et al.
 (2004)  {\em \apj, 616}, L35-L38. 

\refs Stanke T., McCaughrean M. J., and Zinnecker H. (2000) {\em \aap, 355}, 639-650. 

\refs Stahler S. W. (1994) {\em \apj, 422}, 616-620.

      
\refs Stone J. M. and Hardee P. E. (2000) {\em \apj, 540}, 192-210.

\refs Suttner G., Smith M. D., Yorke H. W., and Zinnecker H. (1997) {\em \aap, 318}, 595-607.


\refs Tafalla M. and Myers P. C. (1997) {\em \apj, 491}, 653-662.

\refs Tafalla M., Bachiller R., Wright M. C. H., and Welch W. J. (1997) {\em \apj, 474}, 329-345.

\refs Tafalla M., Santiago J., Johnstone D., and Bachiller R. (2004)
      {\em \aap, 423}, L21-L24.

\refs Tafoya D., G\'omez Y., and Rodr\'{\i}guez L. F. (2004) {\em
      \apj, 610}, 827-834. 

\refs Takakuwa S., Ohashi N., and Hirano N. (2003) {\em \apj, 590}, 932-943.



\refs Torrelles J. M., Patel N. A., Anglada G., G\'omez
      J. F., Ho P. T. P., et al.
      (2003) {\em \apj, 598}, L115-L119.

\refs van Dishoeck E. F. (2004) {\em Ann. Rev. Astron. Astrophys., 42}, 119-167.

\refs van der Tak F. F. S. and Menten K. M. (2005) {\em \aap, 437}, 947-956.

\refs van der Tak F. F. S., Boonman A. M. S., Braakman R., and  van
Dishoeck E. F. (2003) {\em \aap, 412}, 133-145.

\refs Velusamy T. and Langer W. D. (1998) {\em Nature, 392}, 685-687.

\refs Viti S. and  Williams D. A. (1999) {\em \mnras, 310}, 517-526.

\refs Viti S., Codella C., Benedettini M., and Bachiller R. (2004) 
    {\em \mnras, 350}, 1029-1037.

\refs V\"olker R., Smith M. D., Suttner G., and Yorke H. W. (1999) {\em \aap, 343}, 953-965.

\refs Wakelam V., Caselli P., Ceccarelli C., Herbst E., and  Castets
A. (2004) {\em \aap, 422}, 159-169.

\refs Wakelam V., Ceccarelli C., Castets A., Lefloch B., Loinard L., et al.
(2005) {\em \aap, 437}, 149-158.

\refs Walker C. K., Carlstrom J. E., and Bieging J. H. (1993) {\em
      \apj, 402}, 655-666.

\refs Watson C., Zweibel E. G., Heitsch  F., and Churchwell E. (2004) {\em \apj, 608}, 274-281.

\refs White  G. J. and Fridlund C. V. M. (1992) {\em \aap, 266}, 452-456.

\refs Wilkin F. P. (1996) {\em \apj, 459}, L31-L34.

\refs Williams J. P., Plambeck R. L., and Heyer M. H. (2003) {\em \apj, 591}, 1025-1033.

\refs Wiseman J., Wootten A., Zinnecker H., and McCaughrean M. (2001) {\em \apj, 550}, L87-L90.
      
\refs Wu Y., Wei Y., Zhao M., Shi Y., Yu W., Qin S., and Huang 
      M. (2004) {\em \aap, 426}, 503-515.

\refs Yamashita T., Suzuki H., Kaifu N., Tamura M., Mountain
      C. M., and Moore T. J. T. (1989) {\em \apj, 347}, 894-900.

\refs Yokogawa S., Kitamura Y., Momose M., and Kawabe R. (2003) {\em \apj, 595}, 266-278.

\refs Yorke H. W. and Sonnhalter C. (2002)  {\em \apj, 569}, 846-862.

\refs Yu K. C., Billawala Y., and Bally J. (1999) {\em \aj, 118}, 2940-2961.

\refs Yu K. C., Billawala Y., Smith M. D., Bally J., and Butner H. M. (2000)
{\em  \aj, 120}, 1974-2006.

\refs Zhang Q. and Zheng X. (1997) {\em \apj, 474}, 719-723.

\refs Zhang Q., Hunter T. R., Sridharan T. K., and Ho T. P. T. (2002) {\em \apj, 566}, 982-992.

\refs Zhang Q., Hunter T. R., Brand J., Sridharan T. K., Cesaroni R., et al.
(2005) {\em \apj, 625}, 864-882.


\end{document}